\documentclass[useAMS,usenatbib]{mn2e}

\title[Clumpy and fractal shocks, and the generation of a velocity dispersion in molecular clouds]
{Clumpy and fractal shocks, and the generation of a velocity dispersion in molecular clouds}
\author[C. L. Dobbs and I. A. Bonnell]
{C. L. Dobbs$^{1,}$$^2$\thanks{E-mail:
dobbs@astro.ex.ac.uk} and 
I. A. Bonnell$^1$ \\
$^1$School of Physics and Astronomy, University of St Andrews, 
North Haugh, St Andrews, Fife, KY16 9SS\\
$^2$School of Physics, University of Exeter, Stocker Road, EX4 4QL\\}
\usepackage{amssymb}
\usepackage{amsmath}
\usepackage{psfig}

\begin{document}

\date{\today}

\pagerange{\pageref{firstpage}--\pageref{lastpage}} \pubyear{0000}

\maketitle

\label{firstpage}

\begin{abstract}
We present an alternative explanation for the nature of turbulence in 
molecular clouds. Often associated with classical models of turbulence, 
we instead interpret the observed gas dynamics as random motions, induced 
when clumpy gas is subject to a shock. 
From simulations of shocks, we show that 
a supersonic velocity dispersion occurs in the
shocked gas provided the initial distribution of gas is sufficiently 
non-uniform. 
We investigate the velocity size-scale relation $\sigma \propto r^{\alpha}$ for
simulations of clumpy and fractal gas, and show that clumpy shocks can produce 
realistic velocity size-scale relations with mean $\alpha 
\thicksim 0.5$.
For a fractal distribution, with a fractal dimension of 2.2 similar to what is 
observed in the ISM,  we find $\sigma \propto r^{0.4}$.
The form of the velocity size-scale relation can be understood as due to mass
loading, i.e. the post-shock velocity of the gas is determined by the amount of 
mass encountered as the gas enters the shock. We support this hypothesis with
analytical calculations of the velocity dispersion relation for different 
initial distributions.
 A prediction of this model is that the line-of sight velocity dispersion
should depend on the angle at which the shocked gas is viewed.  
\end{abstract}


\section{Introduction}
Molecular clouds are known to exhibit supersonic chaotic dynamics
\citep{Heyer2004,Ossenkopf2002,Falgarone1990,Perault1985,
Hayashi1989,Larson1981}, which are thought to
control star formation and determine the properties of protostellar cores 
\citep{MacLow2004}. Although referred to as 
'turbulence', the origin and nature of these motions are not fully understood. 
The most general
definition of ISM turbulence is simply that the gas exhibits random 
motions on many scales \citep{MacLow2004}. However there is a 
consistent correlation observed in molecular clouds between  
the velocity dispersion and size scale (the 
\citet{Larson1981} relation), approximately
$\sigma \propto r^{0.5}$ (e.g. \citet{Myers1983,Solomon1987,Brunt2003}). 
This has invoked many comparisons between interstellar turbulence and
classical turbulence, e.g. Kolmorogov incompressible turbulence 
\citep{Passot1988,Falgarone1990} ($\sigma \propto L^{0.33}$);
Burger's shock dominated turbulence \citep{Scalo1998} ($\sigma \propto
L^{0.5}$);
and the She-Leveque model for incompressible turbulence 
\citep{She1994,Boldyrev2002} 
($\sigma \propto L^{0.42}$ \citep{Bold2002}). 

The possible sources of turbulence can be summarised as follows: 
gravitational, magnetic or hydromagnetic instabilities; 
galactic rotation, through magneto-rotational instabilities, 
shocks in spiral arms or collisions of clouds on different epicyclic orbits; 
stellar feedback via supernovae, stellar winds and HII regions.  
Recent simulations have 
indicated that turbulence induced by a large scale driving force 
(e.g. large scale flows from supernovae or galactic rotation)
is more 
consistent with observed molecular cloud structures \citep{Brunt2003,
Klessen2001}. Supernovae have been shown to produce sufficient energy to
generate the velocity dispersions observed \citep{MacLow2004}. 
However observations of turbulent velocities in regions which do
not contain massive star formation suggests that other mechanisms, such as 
magneto-rotational instabilities
\citep{Piontek2005,Sellwood1999} and colliding flows \citep{Ball1999} must 
also be important. 
Interestingly, recent observations have 
suggested that the elongations of molecular clouds are more compatible with 
galactic rotation models rather than stellar feedback \citep{Koda2006}.

Galactic disk simulations have investigated gravity driven turbulence
\citep{Wada2002}, stellar feedback \citep{Wada2001,Avillez2005,
Dib2006} and the influence of spiral density waves on ISM dynamics 
\citep{Dobbs2006}.
Analytical results also indicate that vorticity is generated
in centrally condensed clouds subject to galactic shocks \citep{Kornreich2000},
and the induced velocities follow the observed velocity size-scale relation.  
Previous numerical work on colliding flows showed that density and velocity
perturbations occur even in uniform flows subject to cooling instabilities 
\citep{Heitsch2005}, although a velocity length scale correlation was not 
investigated.
Simulations of clumpy flows have also indicated that a Salpeter type clump mass spectrum
can be reproduced \citep{Clark2006}.

Spiral shocks have also been proposed to explain the dynamics of molecular
clouds \citep{Bonnell2006,Zhang2001}.
\citet{Bonnell2006} model giant molecular cloud formation as gas
passes through a clumpy spiral shock. The dynamics of the 
molecular clouds are determined on all scales 
simultaneously as the clouds form and the induced velocity
dispersion size scale relation is consistent with observations. 
This can account for the observed velocity dispersions that are found even in
regions devoid of massive stars.
Furthermore, there is no need for a continuous 
driving mechanism as the time for the decay of these velocities is proposed to 
be of similar magnitude to the cloud lifetime. 
 
In this paper we investigate the velocity size relation in shock tests with 
uniform, clumpy and fractal distributions of isothermal gas.
The clumpy and fractal distributions are chosen to reflect the highly 
structured nature of the ISM \citep{Cox2005,Elmegreen2004,Dickey1990,
Perault1985}. 
We concentrate on modelling 
the passage of gas through a spiral arm by using a linear sinusoidal potential,
although these results would apply generally to shocks between colliding flows.
We show that random velocities 
induced in non-uniform shocks display a velocity size relation similar to
those observed and provide simple analytical analysis alongside the results of 
our simulations.
Thus the 'turbulence' in our results describes random motions of
the gas and does not correspond to any theories of classical turbulence.

\section{Calculations}
We use the 3D smoothed particle hydrodynamics (SPH) code based on the version by
Benz \citep{Benz1990}. The smoothing length is allowed to vary with space and
time, with the constraint that the typical number of neighbours for each particle 
is kept near $N_{neigh} \thicksim50$.  
Artificial viscosity is included with either the standard parameters $\alpha=1$
and $\beta=2$ \citep{Monaghan1985,Monaghan1992} or $\alpha=2$ and $\beta=4$.

\subsection{Initial conditions}
We investigate uniform, clumpy and multi-scale shocks, starting with 3D 
shock tube tests first before considering gas subject to an external potential.
In all calculations, the gas is isothermal, non
self-gravitating and there are no magnetic fields. These calculations are 
dimensionless, and are characterized by the Mach number of the shock and the 
initial density distribution of the gas. 
In all calculations, the particles are allocated velocities in
the $x$ direction only and, except for the oblique shocks (Section~3.2), 
the gas shocks in the $yz$ plane.

The parameters used in the different simulations are shown in Table~1.
We investigated 4 distributions of gas - uniform, homogeneous spherical clumps 
in pressure equilibrium, spherical clumps of different radii/density
and fractal distributions.
The filling factors for each distribution are calculated by overlaying a 3D grid
on each distribution and determining the porosity. The filling factor is given
by $F=N_{full}/N_{total}$ where $N_{full}$ is the number of cells containing at
least 1 particle and $N_{total}$ the total number of cells. 
We take a 32$^3$ grid, so the mesh resolution is equivalent to a couple of
smoothing lengths for the uniform case. 
This ensures that 
each cell contains at least 1 particle for the
uniform distribution so the filling factor is 100\%.   
The maximum initial scale length corresponds to the range of $x$
values for the particles in the initial distribution. 
\begin{table}
\centering
\begin{tabular}{|c|c|c|c|c|c|}
\hline
Test & Distribution & L & F (\%) & 
${\cal M}$\\ \hline
ST & Uniform & 4 & 100 & 10 \\
ST & Clumps ($r_{cl}$=0.1) & 4 & 27 & 20 \\
ST & Clumps ($r_{cl}$=0.2) & 4 & 50 & 20 \\
ST & Uniform & 4  & 100 & 20 \\
ST & Clumps ($r_{cl}$=0.1) & 4 & 27 & 40 \\
ST & Clumps ($r_{cl}$=0.2) & 4 & 50 & 40 \\
SP & Uniform & 3 & 100 & 30 \\
SP & Clumps ($r_{cl}$=0.1) & 3 & 12 & 30 \\
SP & Clumps ($r_{cl}$=0.2) & 3 & 23 & 30 \\
SP & Clumps ($r_{cl}$=0.4, 0.1, 0.04) & 3 & 32 & 30 \\
SP & 2.2 D Fractal & 3 & 10 & 30 \\ 
SP & 2.7 D Fractal & 3 & 23 & 30 \\ \hline
\end{tabular}
\caption{Table showing the different runs performed and the distributions of gas used 
for initial conditions. ST and SP refer to shock tube and
sinusoidal potential tests, L is the maximum initial length scale, 
F the volume filling factor and ${\cal M}$ the Mach number of the shock.
All simulations used $2\times10^5$ particles except for the
fractal distributions which required either $2\times10^6$ or 
$3\times10^6$ particles.}
\end{table}

\subsubsection{Shock tube test}
We first perform 3D shock tube tests with initial distributions of uniform and 
clumpy gas to model colliding flows. For both distributions, particles are 
placed within a cuboid 
of dimensions $-2<x<2$, $-1<y<1$ and $-1<z<1$. 
To produce a clumpy shock we distribute the particles in
uniform density spheres within these length scales.
The clumps are a constant temperature, 
confined by either an external pressure field or a
hotter diffuse phase.
The clumps are initially allowed to 
settle into equilibrium before the simulation is carried out. 
To produce clumps of different radii, the
external pressure (or pressure of the diffuse phase) can be increased or 
decreased. 
We then assign particles a velocity of $v_0=10$ $c_s$ for $x\le0$ and
$v_0=-10$ $c_s$ for $x>0$. For the uniform distribution, this produces two
approximately Mach 10 shocks, and multiple shocks of up to Mach 20 for the
clumpy distributions. 
These calculations were also 
repeated using twice the initial velocities and 
all tests use $2\times10^5$ particles.

\subsubsection{Sinusoidal potential}
We mainly consider shocks in gas subject to an
external potential. In this case the gas self shocks, similar to gas
experiencing a stellar potential in a spiral galaxy. A velocity
dispersion relation similar to the observed $\sigma \propto r^{0.5}$ law has
been shown to develop as gas passes through a clumpy spiral shock 
\citep{Bonnell2006}. Here we examine a simplified setup, 
where we can investigate the effect of the shock dynamics on the initial gas 
distribution. Instead of the spiral potential we use a 1D sinusoidal 
potential of the form 
\begin{equation}
\psi=A\cos(k(x+B)).
\end{equation}
where $k$ is the wavenumber and $B$ a length parameter to determine the location
of the minimum.
This is equivalent to the linear passage of gas through sinusoidal spiral arms.
The dimensionless velocity acquired by gas falling from the peak to the base of
the potential is $V_{pot}=\sqrt{2\times A}$.
We tried many different potentials, varying $A$ and $k$, and only
applying the potential once the gas has passed a minimum. However we found that
the results presented here are largely independent of the exact nature of the
potential. The structure of the shock is similar for different potentials for
a given initial distribution. The relative strength of the shock determines the 
magnitude of the velocity dispersion, whilst the initial distribution determines
the velocity size scaling law. For the simulations
presented here, we took $k=\pi/4$, $A=100$ and $B=2$ to produce a minimum at 
2 and maxmima at -2 and 6.

We allocate particles a velocity of 50 $ c_s$ in the $x$ direction and zero
velocity in the $y$ and $z$ directions, which for the simulations described here,
leads to a shock of Mach number $\approx 30$.
We set up a distribution of spherical clumps in pressure
equilibrium in the same way as described for the shock tube tests. Where a hot
diffuse phase is used to supply an external pressure, the hot phase is distributed
with the same number of particles (2$ \times 10^5$), but 1/10 of the mass of the
cold phase. 

We also test a distribution with clumps of different size-scales and densities,
giving structure on a range of scales. The clumps have initial diameters of 
0.4, 0.1 and 0.04, with clumps of smaller diameter placed inside larger clumps.
For the uniform and clumpy distributions, particles are positioned within a
cuboid of dimensions $-1.5<x<1.5$, $-1<y<1$ and $-1<z<1$.
We also investigate shocks with an initial fractal distribution, following 
the method described in \citet{Elmegreen1997} to generate fractals. The
algorithm includes 3 parameters, an intrinsic length scale $L$, the number of
hierarchical levels, $H$, and the number of points in each level, $N$. The
dimension of the fractal is $D=logN/logL$ and the number of points is $N^H$. 
We generate a 2.2 D fractal, with $L=2.1$, $N=5$ and $H=9$ requiring 
$\approx 2$ million points, and a 2.7 D fractal, with $L=2.5$, $N=12$ and 
$H=6$ requiring $\approx 3$ million points.
We then scale the $x,y,z$
coordinates (equally) of each fractal to fit inside a cube of dimensions
$-1.5<x<1.5$, $-1.5<y<1.5$ and $-1.5<z<1.5$.  
Observations estimate the interstellar fractal dimension as D=2.3
\citep{Elmegreen1996}.    
 
Since for the distributions with fractals or different size clumps the gas
exhibits different densities and pressures, constant pressure boundaries, or an
intervening diffuse phase, are no
longer appropriate. Instead we apply a pressure switch in
order that only gas in the shocked region is subject to pressure forces.
The gas experiences pressure only when $div(v)\le0$ i.e. compression of the gas
is occurring. This enables structure on all scales to be maintained in the gas
distribution before gas reaches the shock. Tests for the uniform density clumps 
showed this method produced similar results compared to when constant pressure 
boundaries were applied.
 
\section{Results}
Column density plots for different simulations are shown in Fig.~1,2,3 (all
for the sinusoidal potential tests).
The uniform shock in Fig.~1 shows a smooth shocked region of approximately
constant density and width. By contrast the clumpy shock (Fig.~1,
middle) shows a much broader shocked region of non-uniform density.
The shock contains more
structure and appears more similar to simulations of turbulence. 
In Fig.~1 (middle), an external pressure field is applied to maintain the clumps
in pressure equilibrium. Fig.~1 (lower) shows a shock for similar size clumps,
where the clumps are instead surrounded by hotter gas. In this case, the hot gas
takes the same numbers of particles as the cold gas, but 1/10 of the mass.
The ratio of the densities of cold to hot gas is $\sim 30$, which is similar to the
ratio of densities of the cold and warm neutral components of the ISM \citep{Cox2005}.
The structure of the cold
gas in the shock 
is very similar whether the clumps are in equilibrium from external
pressure boundaries, or hot gas, indicating that the hot gas has little effect
on the gas dynamics.
 
The distribution of different size clumps shows similar morphology, 
although more
smaller scale structure is apparent in the shocked gas (Fig.~2).
The shocked gas of the fractal distribution (Fig.~3) 
shows more filamentary structure compared to the clumpy distributions. 

\begin{figure}
\centerline{\psfig{file=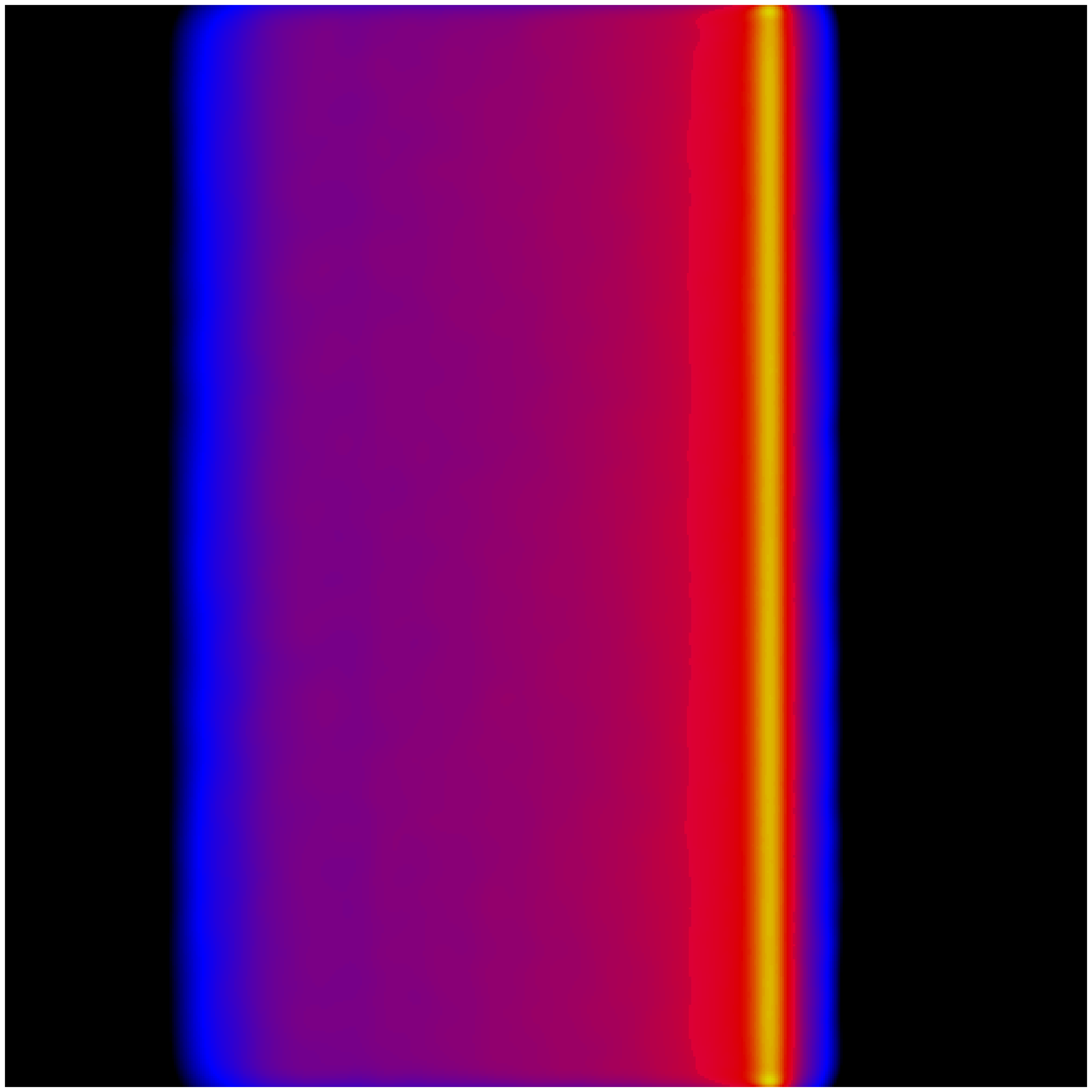,height=1.6in}
\psfig{file=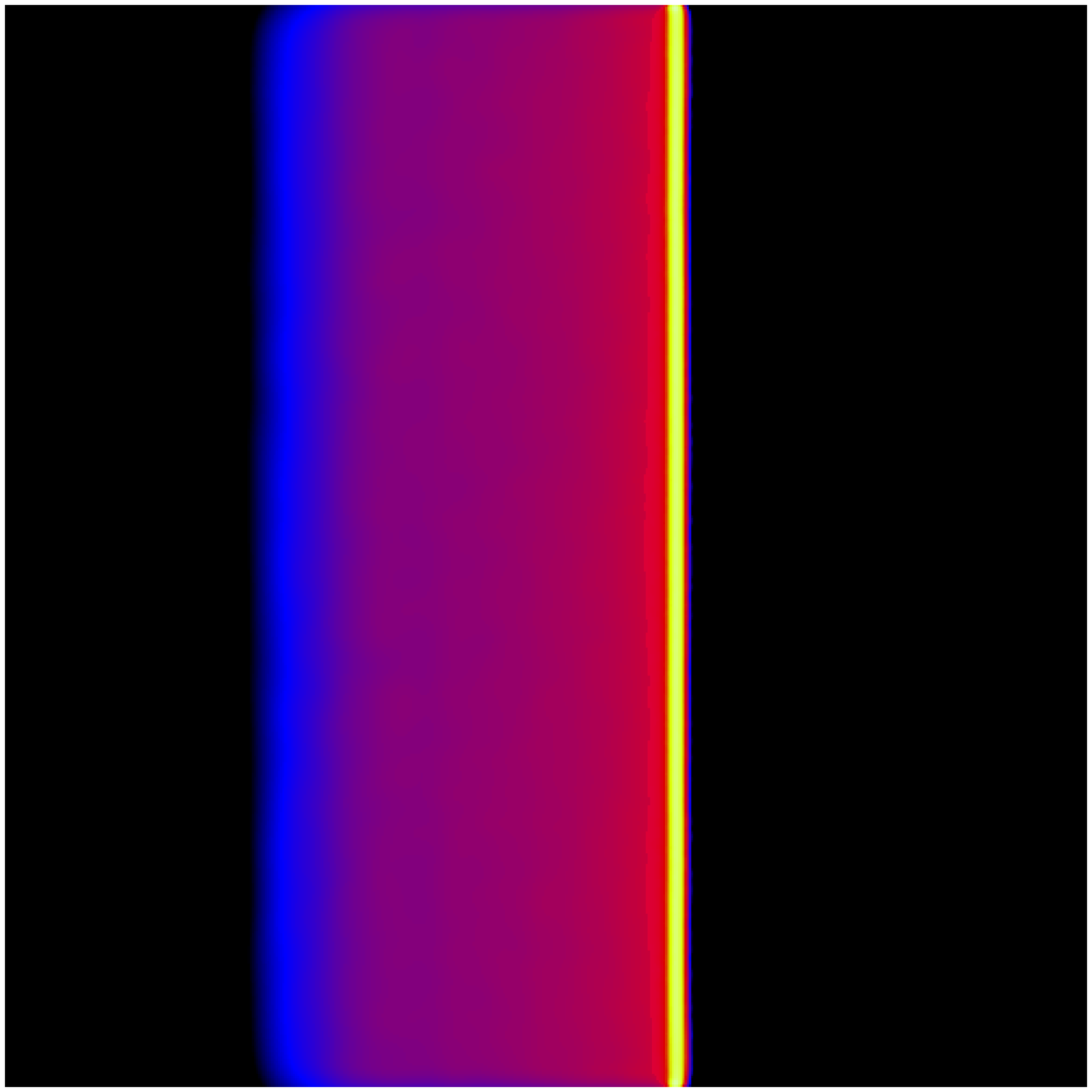,height=1.6in}}
\vspace{0.2pt}
\centerline{\psfig{file=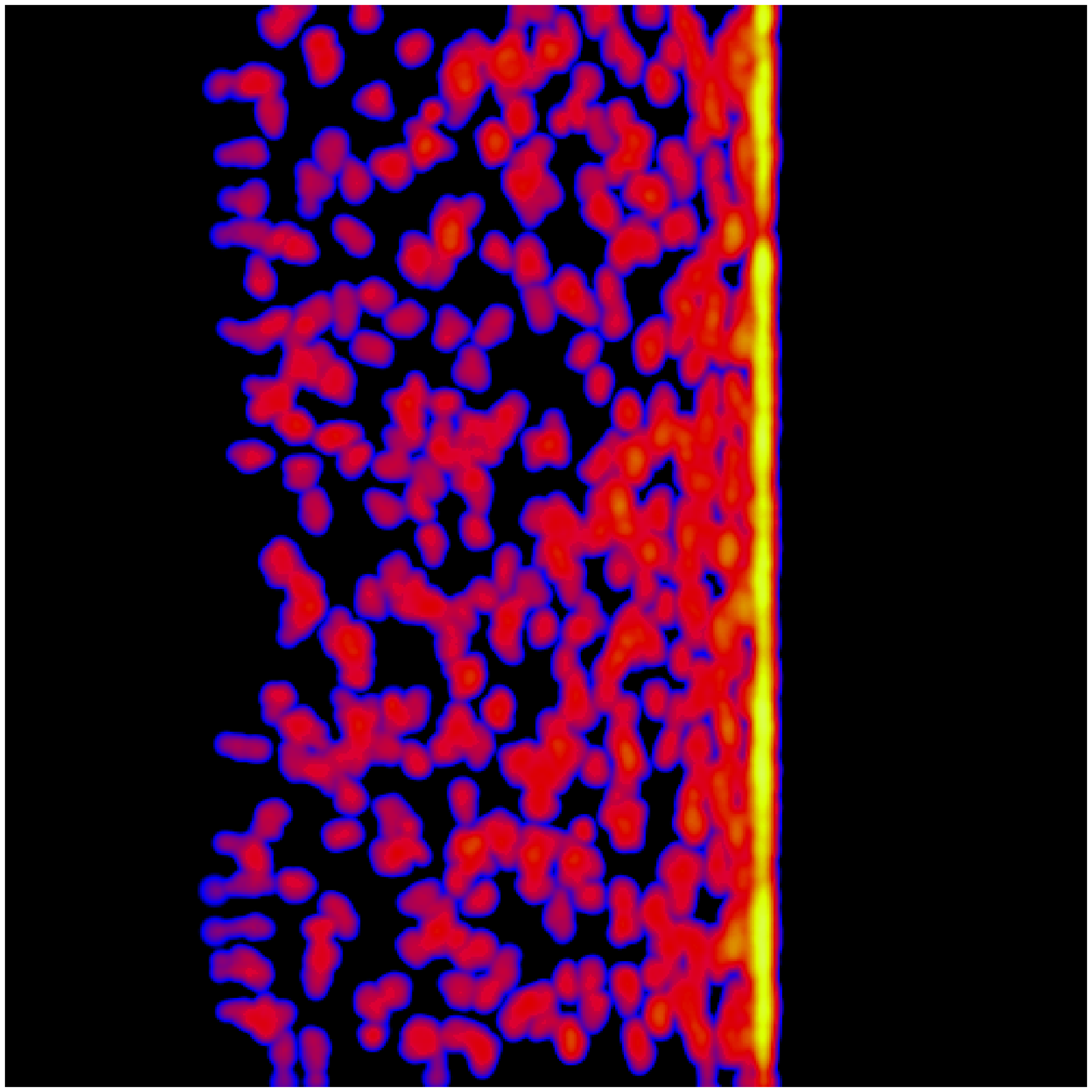,height=1.6in}
\psfig{file=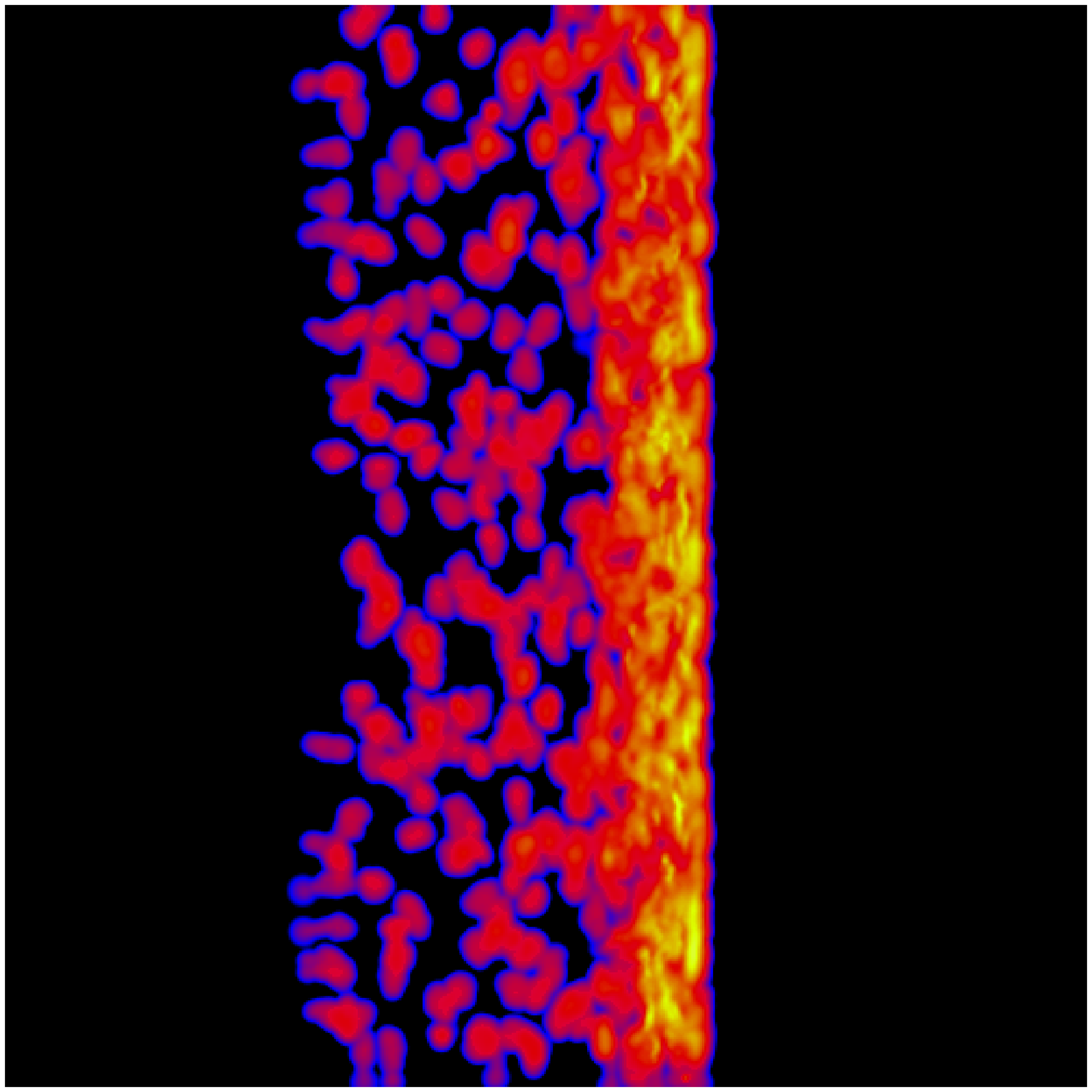,height=1.6in}}
\vspace{0.2pt}
\centerline{\psfig{file=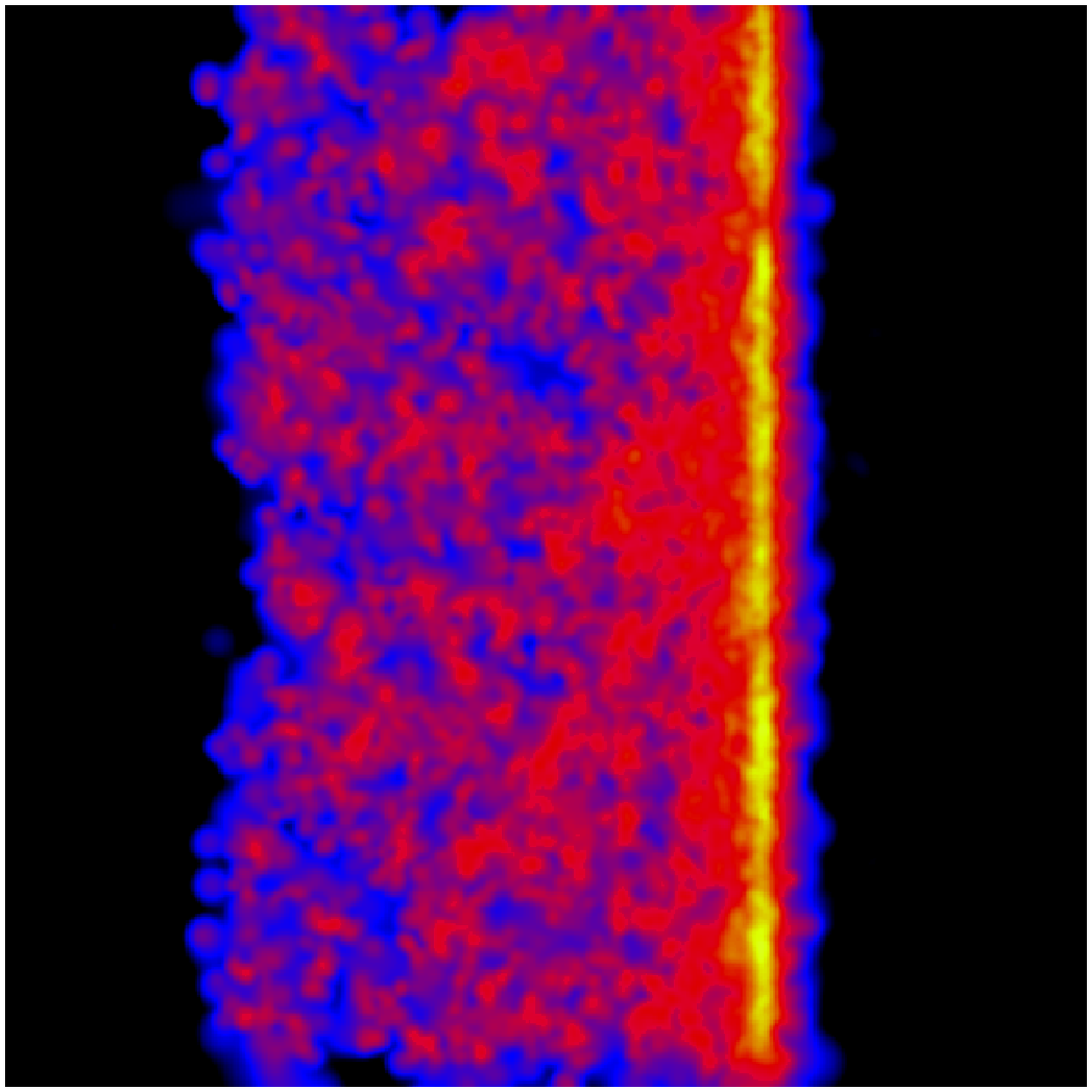,height=1.6in}
\psfig{file=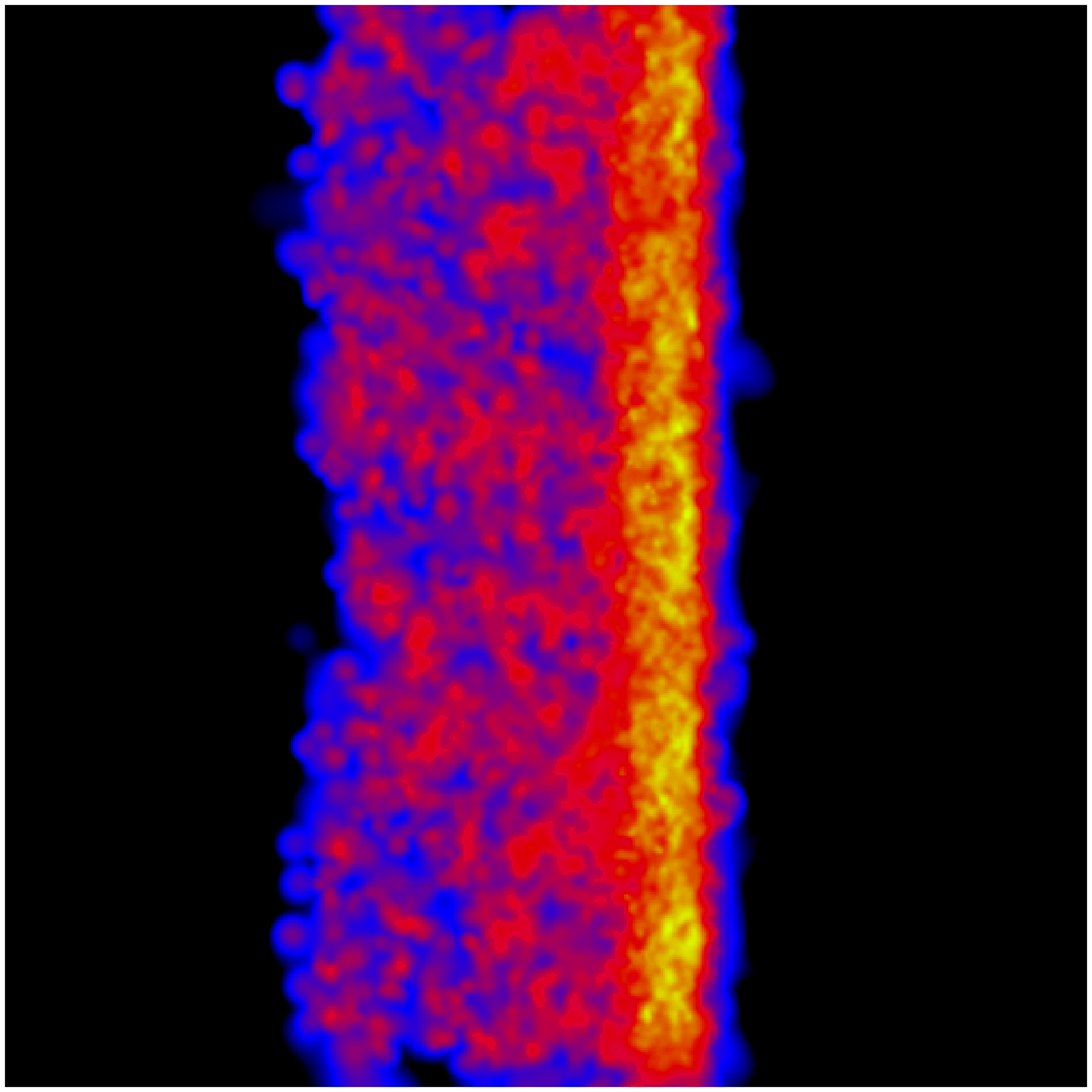,height=1.6in}}
\caption{The column density distribution is shown for linear 
shock tests where gas is passed through a sinusoidal
potential. The initial distribution is uniform (top) and
clumpy (middle, bottom, where the clump radius is 0.1).
The middle panels show the case where the clumps are confined by constant
pressure boundaries, whereas for the bottom panels, the clumps are in
equilibrium with a diffuse phase of gas.
The left panels
($2.8<x<4.8, -1<y<1$) show the gas as it begins to shock (t=0.15) while the 
right panels ($3.2<x<5.2, -1<y<1$) show the
gas when the shock is fully developed (t=0.25). The minimum and maximum of the
potential lie at $x=2$ and $x=6$ respectively. The column density
ranges form 90 to $9\times10^6$ in units of particles per unit area and the
colour scale is logarithmic.}
\end{figure}

\begin{figure}
\centerline{\psfig{file=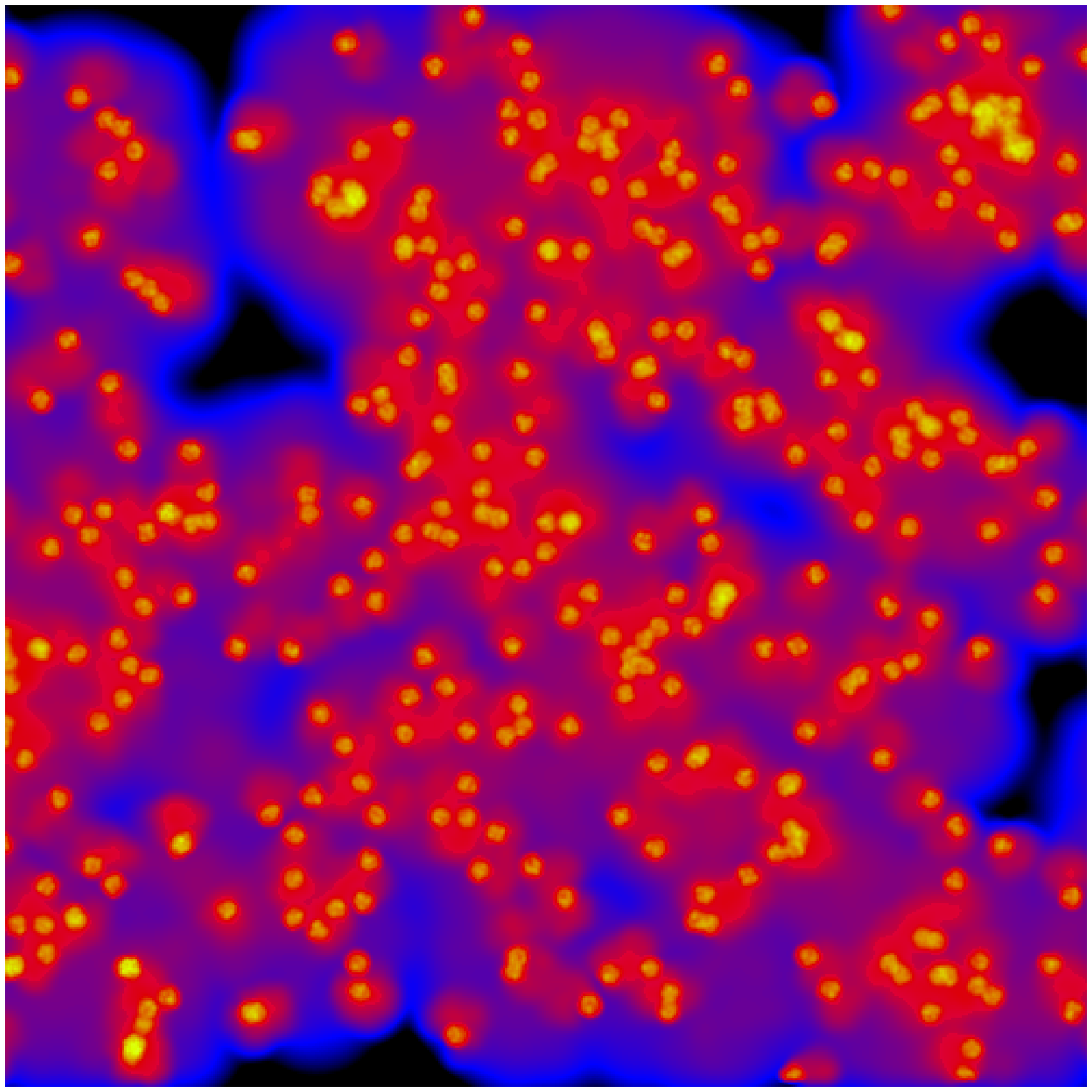,height=1.6in}
\psfig{file=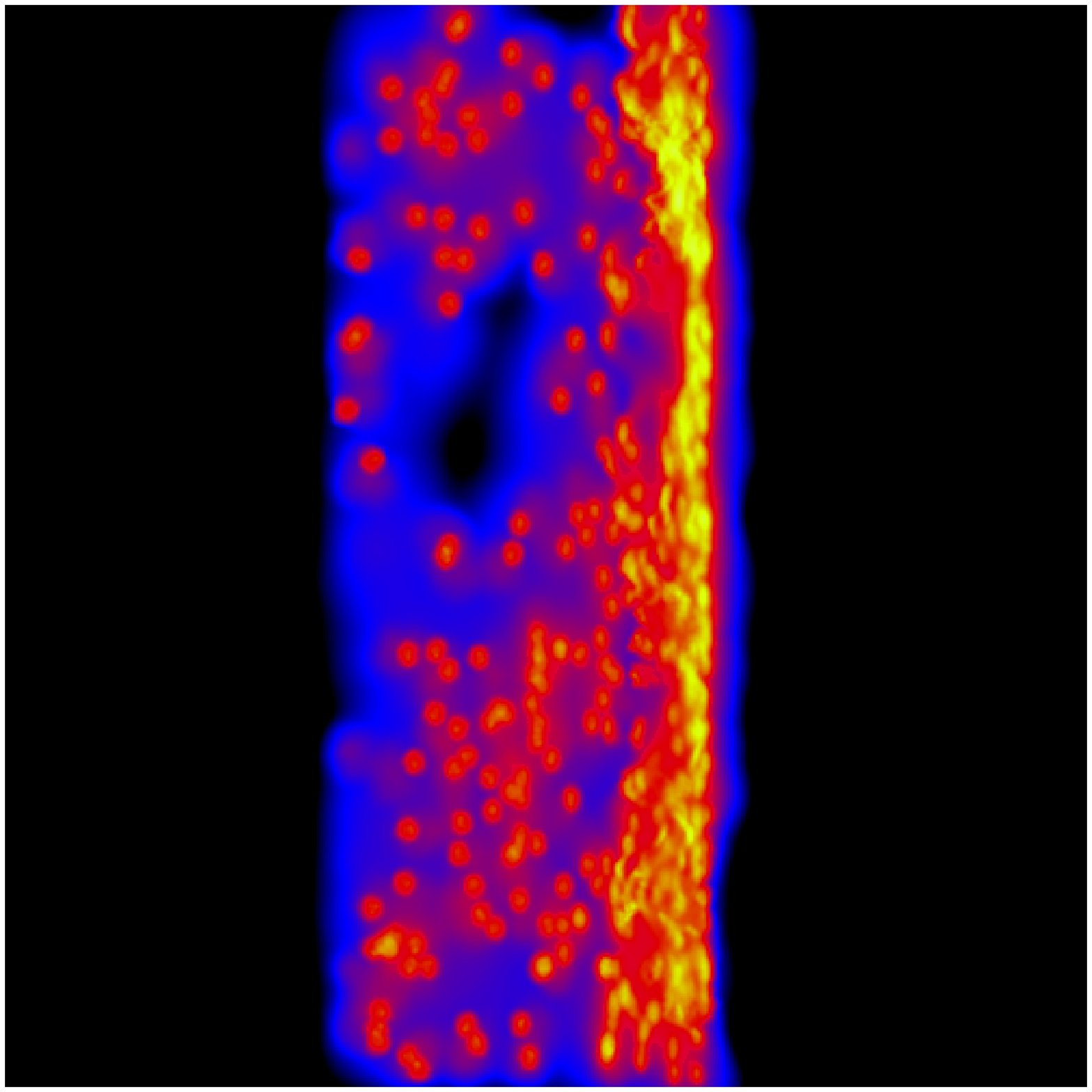,height=1.6in}}
\caption{The column density distribution is shown for a shock test
where the initial distribution consists of different
radii clumps. The gas shocks as it passes through a sinusoidal potential. 
The left panel ($-1<x<1$, $-1<y<1$) shows the initial gas 
distribution (t=0) while the right panel ($3.2<x<5.2$, $-1<y<1$) shows a stage 
(t=0.25) during the shock. The scaling is the same as Fig.~1, and 
the minimum and maximum of the potential lie at $x=2$ and $x=6$.}
\end{figure}

\begin{figure}
\centerline{\psfig{file=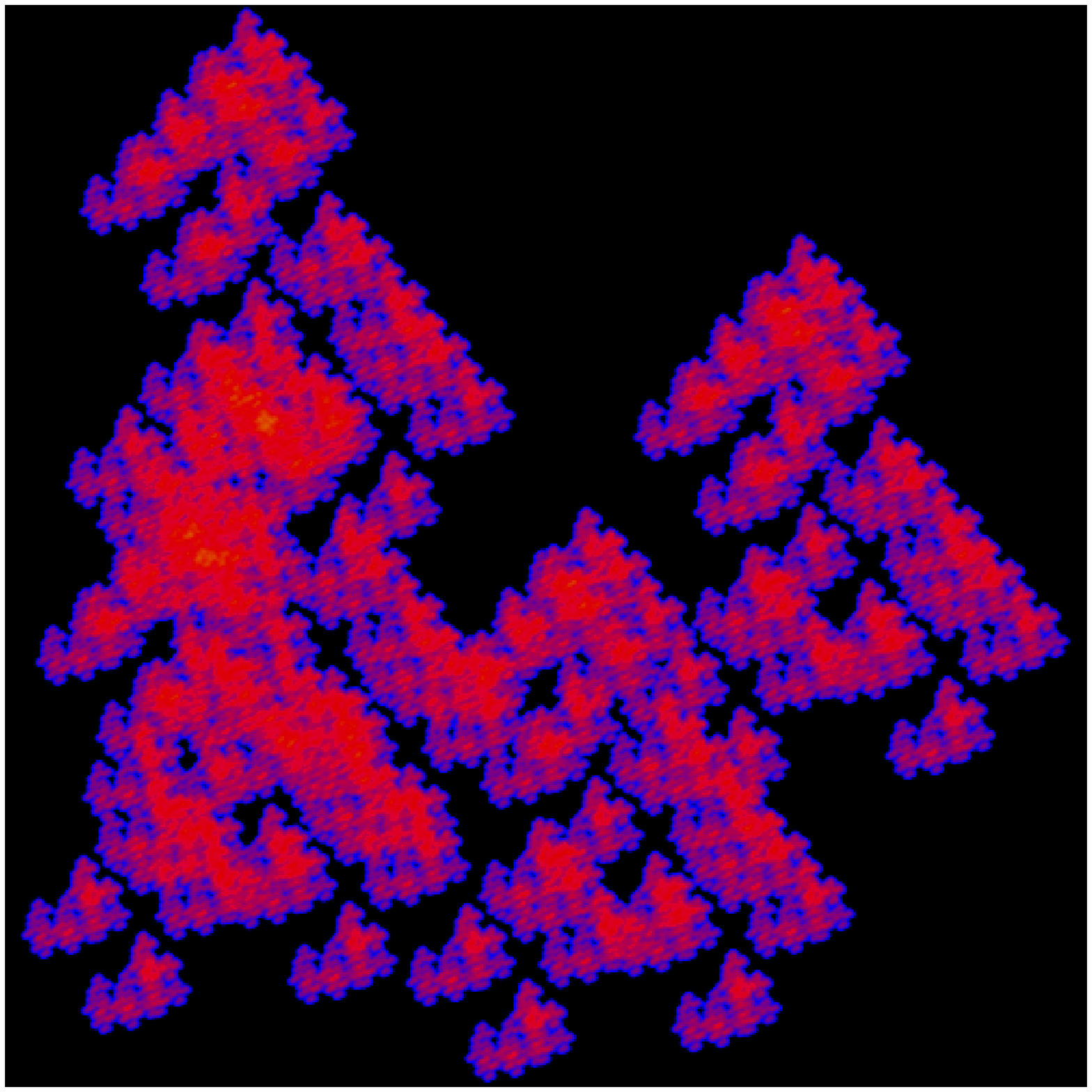,height=1.6in}
\psfig{file=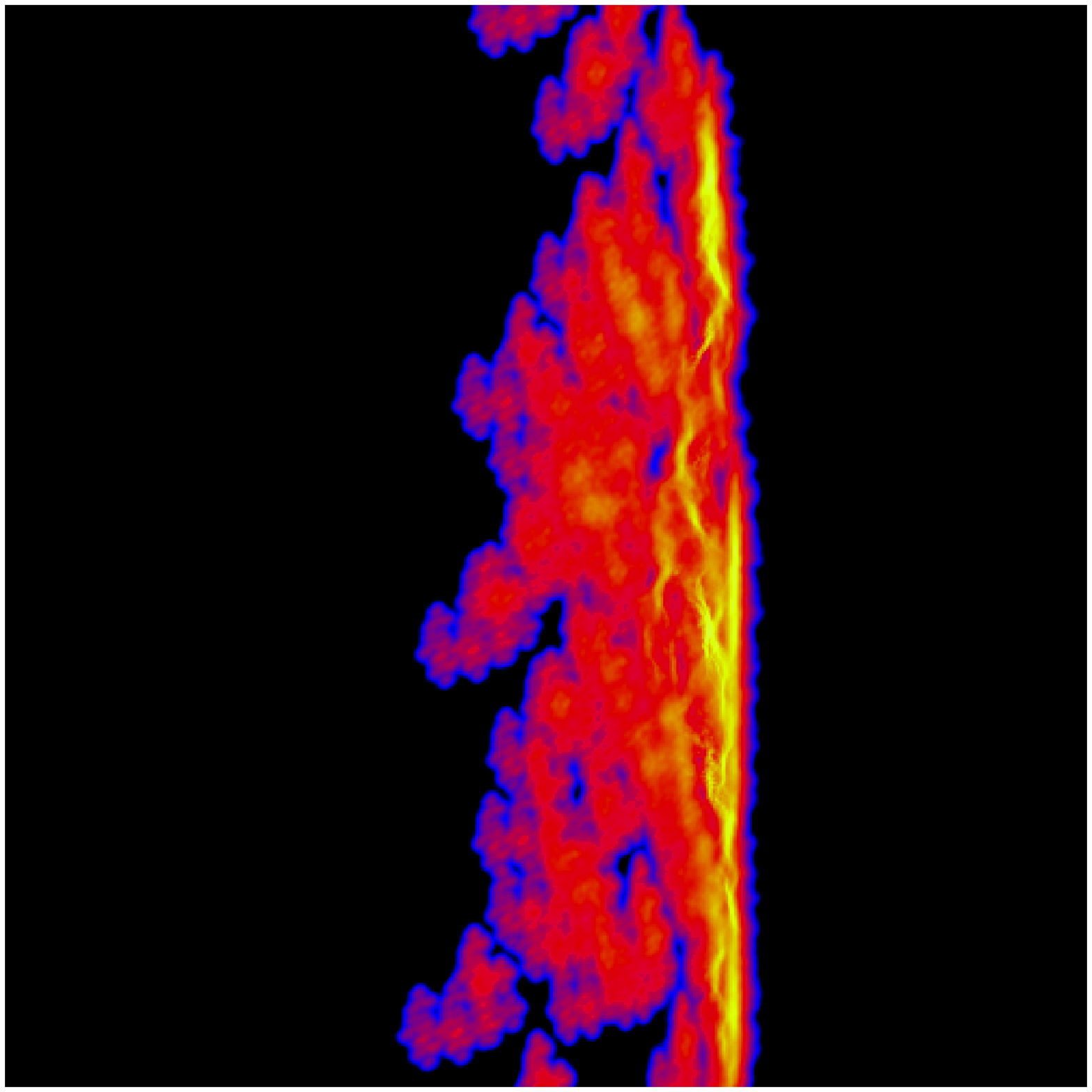,height=1.6in}}
\caption{The column density distribution is shown for a shock test where gas
distributed according to a 2.2D fractal is passed through a sinusoidal
potential. The left panel ($-1.5<x<1.5$, $-1.5<y<1.5$)
shows the initial gas distribution (t=0) 
while the right panel ($3.2<x<5.2$, $-1<y<1$)
shows a stage (t=0.25) during the shock. The scaling is the same as Fig.~1.}
\end{figure}

\subsection{Velocity dispersion}
We now calculate the 1D velocity dispersion of the post-shock gas. We only
consider the $v_x$ velocities, which corresponds to the direction of motion of 
the initial gas, since the velocity dispersion in the $y$ and $z$ directions
are always subsonic. For a given
size, we average the velocity dispersion over numerous regions
of that size scale. The regions are 3 D and chosen to centre on the densest 
particles in the shock.
Only particles with densities greater than the maximum pre-shock density are
considered for calculating the velocity dispersion, thus ensuring we only
include gas in the shock. 
(We find that even for a uniform shock, including the pre and/or post
shock gas will produce a Larson type velocity dispersion size-scale relation,
theoretically and from numerical results.)
We repeat this process for regions of different
size-scale to determine the dependence of the induced velocity dispersion on the
size-scale. 

For the shock tube tests, we initially found that the velocity dispersion was
supersonic, even for the uniform shock. This is due to the inherently clumpy nature 
of SPH which introduces error when calculating the velocity dispersion. 
We therefore increased the viscosity
parameters to $\alpha=2$ and $\beta=4$. This lowered the values of the velocity
dispersion for both the uniform Mach 10 and Mach 20 shocks, although the velocity
dispersion is still supersonic for the Mach 20 shock. The results
presented for the shock tube tests use the higher viscosity parameters, whilst
for the sinusoidal potential tests, there is less noise and the standard
parameters $\alpha=1$ and $\beta=2$ are used. In all cases the higher viscosity
parameters had little effect on the velocity dispersions for the clumpy shocks.
Alternatively, the velocity dispersion can be determined from the SPH smoothed 
velocities:
\begin{equation}
\widehat{\mathbf{v_i}}=\mathbf{v_i}+\sum_{j \ne i} \frac{m_j}{\rho_j} 
\mathbf{v_{ij}} \mathbf{W_{ij}}.
\end{equation}
Since these velocities are smoothed over the neighbouring particles, the velocity
dispersion produces less noise. Consequently the velocity dispersion is lower
(and in all our results subsonic) for the uniform shocks. 
Except for Fig.~6 though, we use the SPH 
velocities, since these are the velocities produced by the code.     

\begin{figure*}
\centerline{\psfig{file=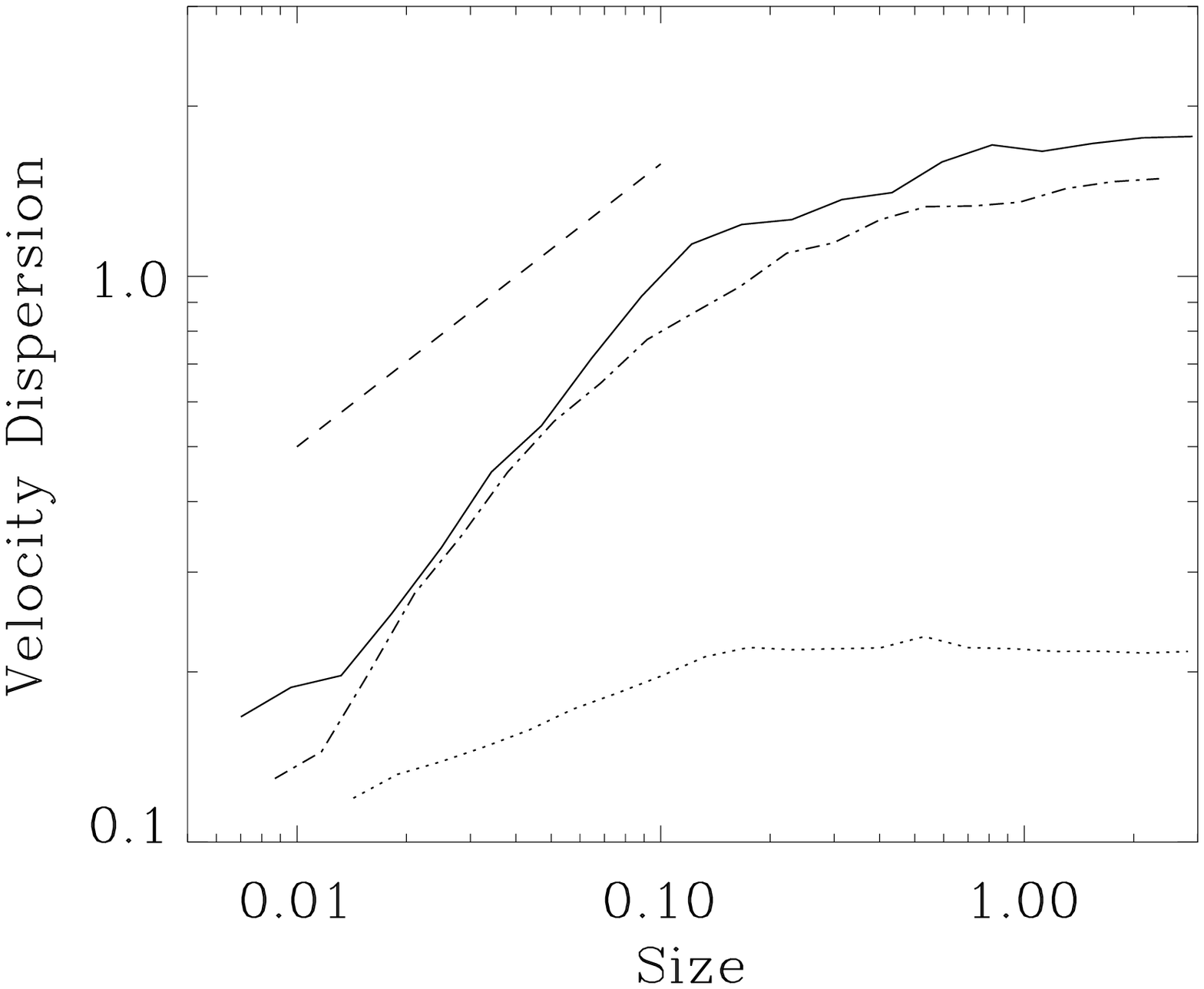,height=2.in}
\psfig{file=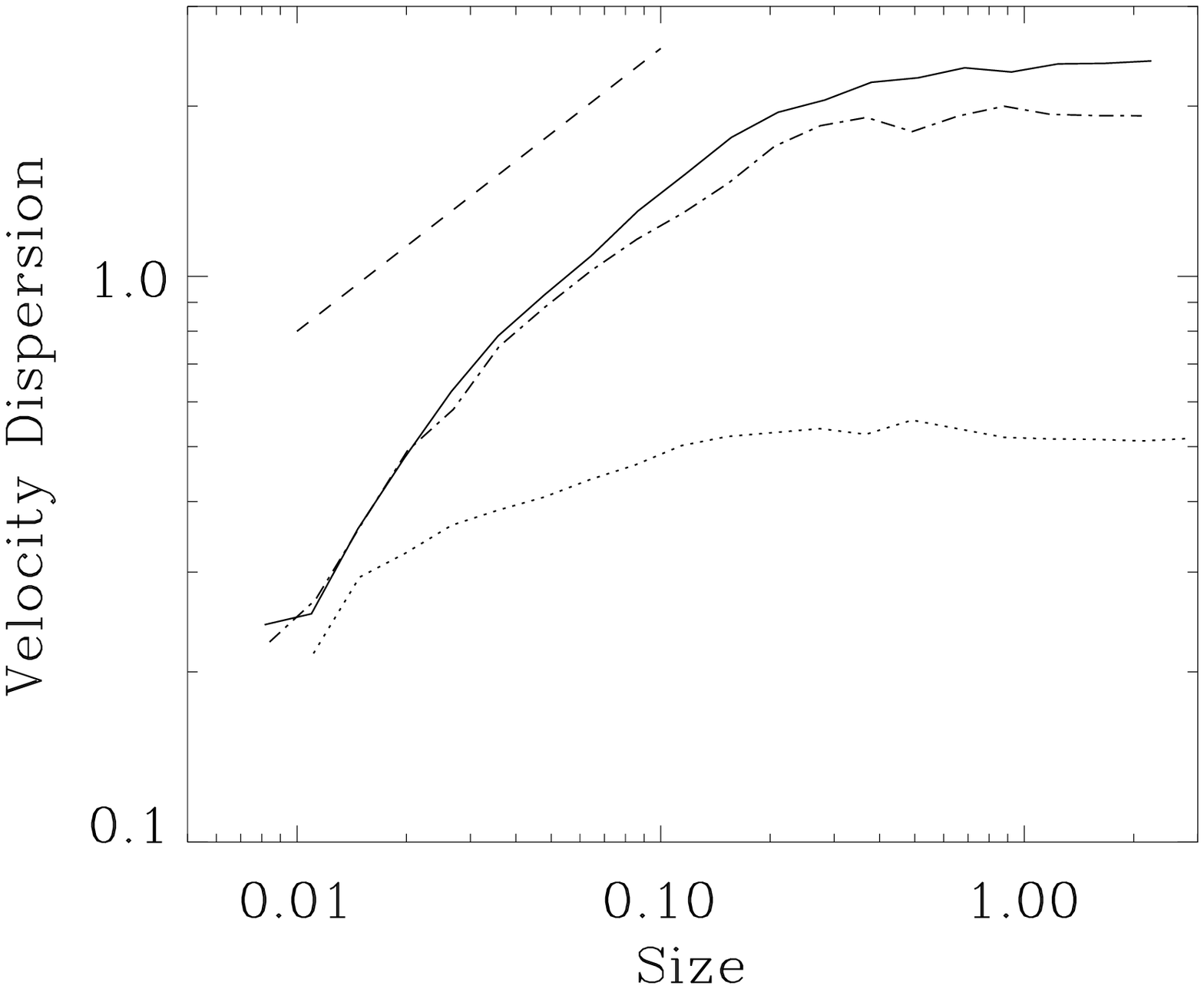,height=2.in}}
\caption{The one-dimensional velocity dispersion of the post-shock gas is plotted
against size-scale for the 3D
shock tube test. The figures show a ${\cal M}$=20 shock (left) and ${\cal M}$=40 shock (right). 
The initial
distributions prior to the shock are uniform (dotted line), clumps of radii
0.1 (solid line) and clumps of radii 0.2 (dash-dot line). External pressure
boundaries are applied to keep the clumps in equilibrium. The dashed line
represents the observed $\sigma \propto r^{0.5}$ relation and the sound speed is 0.3. 
The initial distribution has a maximum spatial extent of 4, and the viscosity
parameters $\alpha=2$ and $\beta=4$ are used.}
\end{figure*}
\begin{figure*}
\centerline{\psfig{file=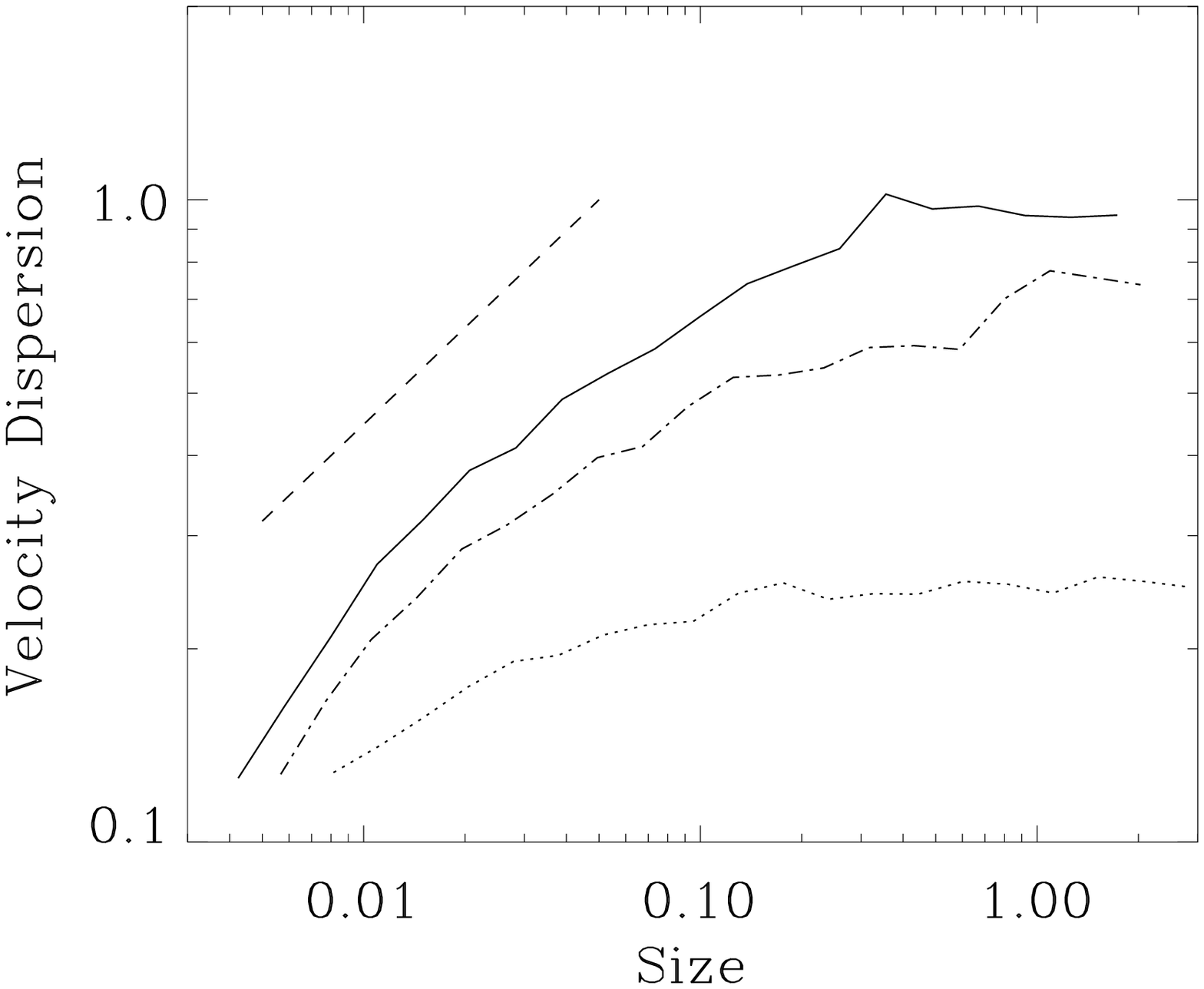,height=2.in}
\psfig{file=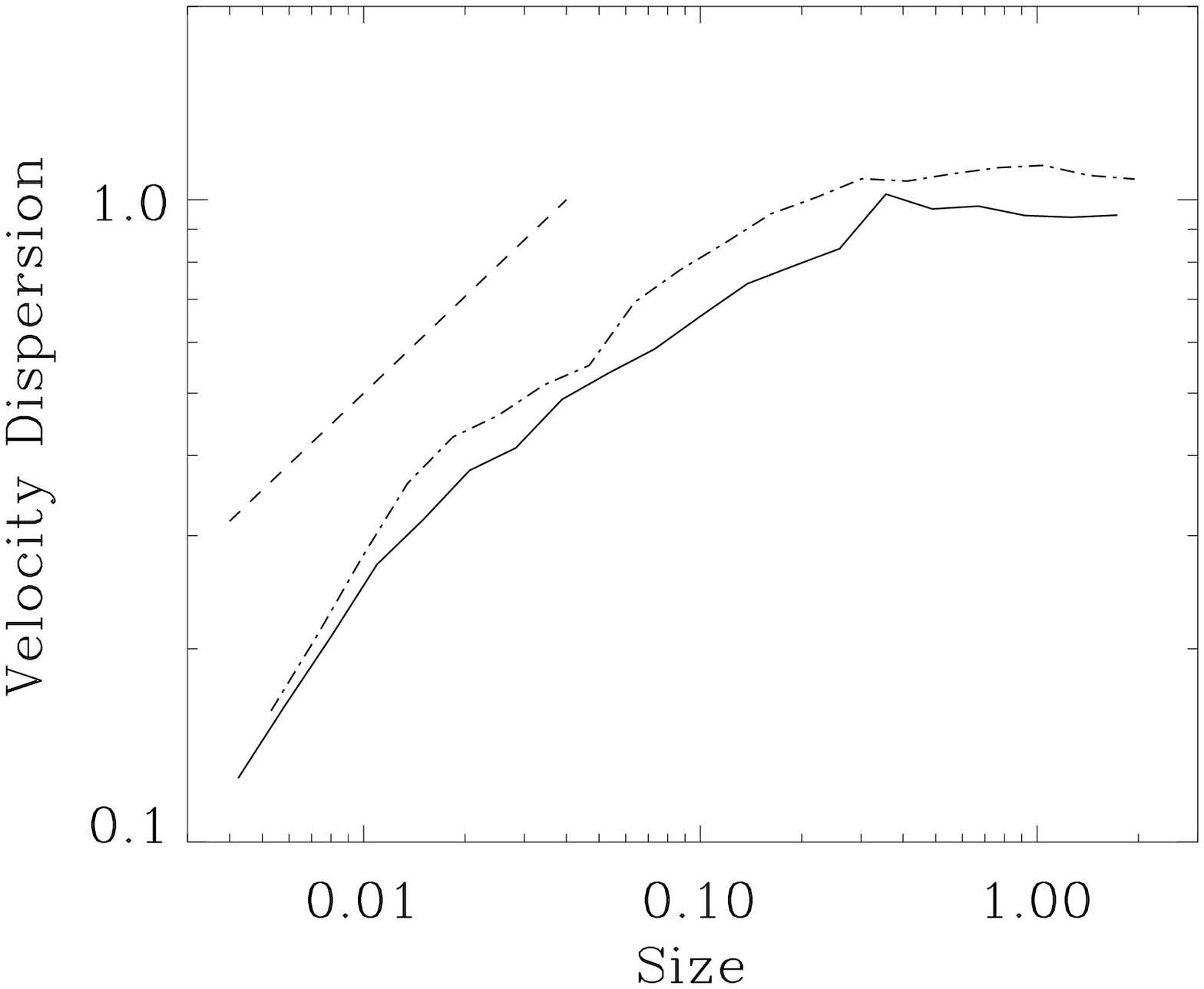,height=2.in}}
\caption{These panels show the one-dimensional velocity dispersion of
the post-shock gas plotted
against size-scale for the sinusoidal potential. The left panel uses the same
initial distributions (with the same key) prior to the shock as Fig.~4, although the
maximum initial length scale is 3 for these distributions.
The right panel compares the velocity dispersion when external pressure
boundaries are applied (solid line) and a diffuse intervening medium is used
(dot dash line). For the right panel, the clump radius is 0.1, so these results
correspond to the column density images in Fig. 1 (middle and bottom).  
The dashed line
shows $\sigma \propto r^{0.5}$ and the sound speed is 0.3.}
\end{figure*}
\begin{figure*}
\centerline{
\parbox{.5\linewidth}{
\centerline{\psfig{file=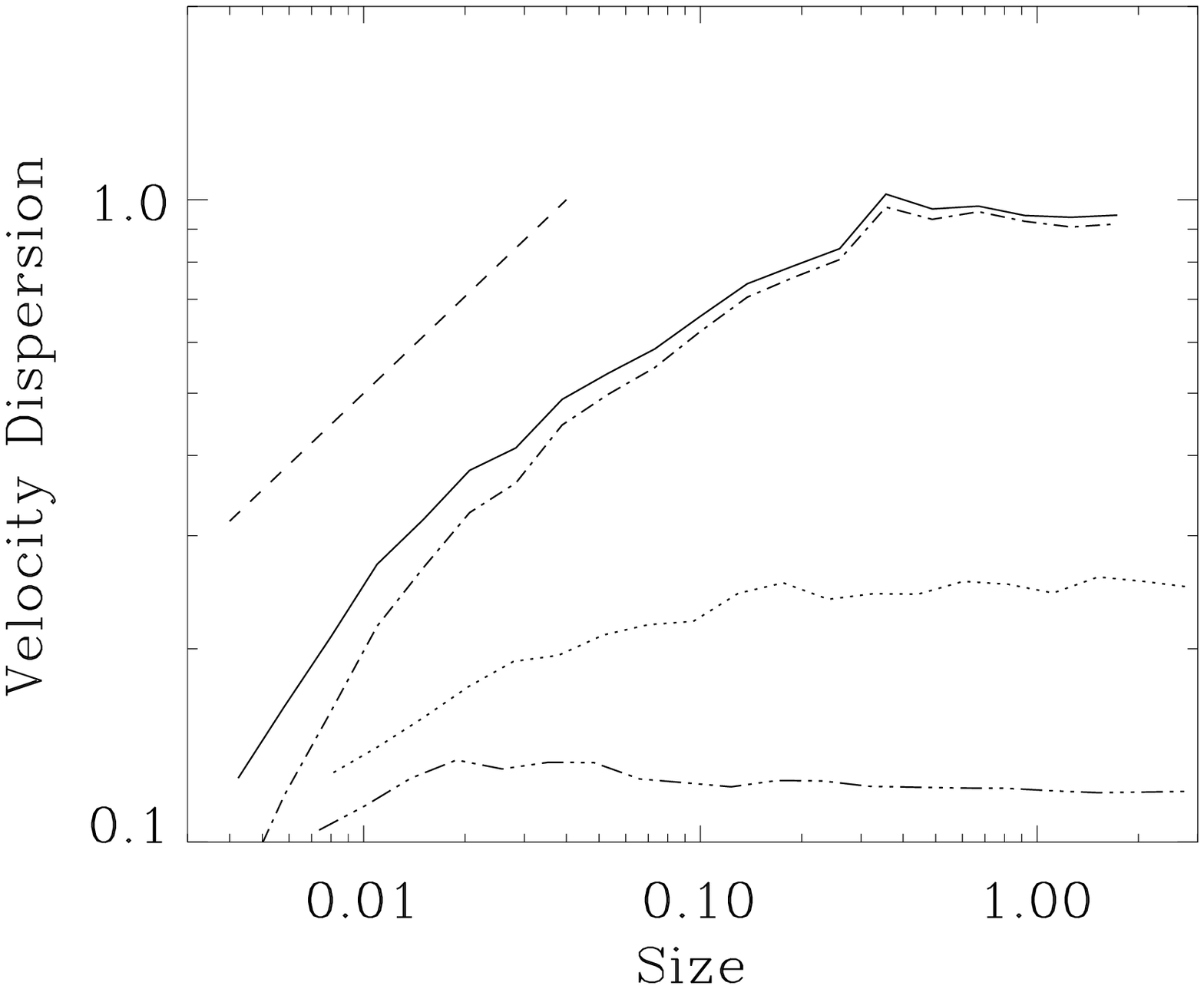,height=2.in}}
\caption{The 1D velocity dispersion relation for is plotted for
post-shock gas using the actual and 
smoothed velocities. The results are shown for 
the r=0.1 clumpy distribution (solid line, actual velocities, dot-dash line, smoothed 
velocities) and the uniform distribution (dotted line, actual velocities, dot-dot-dot dash
line, smoothed velocities), where the gas passes through the sinusoidal potential. 
The dashed line
shows $\sigma \propto r^{0.5}$ and the sound speed is 0.3.}}
\hspace{5pt}
\parbox{.5\linewidth}{
\centerline{\psfig{file=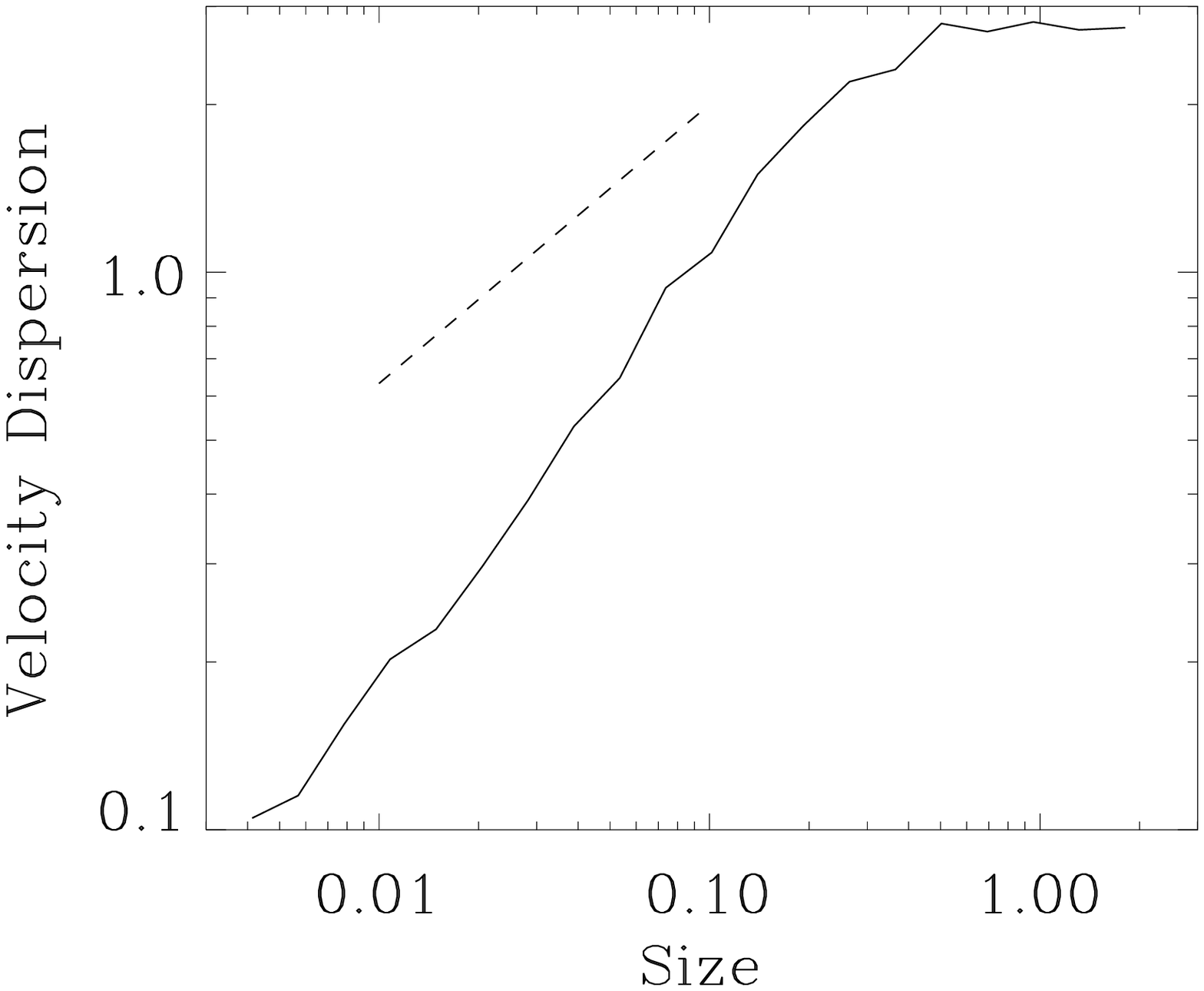,height=2.in}}
\caption{{The 1D velocity dispersion dependence on size-scale is plotted
for post-shock gas using an initial distribution of different size clumps. 
The maximum initial length scale is 3 and the gas shocks when passed through a
sinusoidal potential.
The dashed line
show $\sigma \propto r^{0.5}$ and the sound speed is 0.3.}}}}
\end{figure*}

We plot the velocity dispersion size-scale relation for each of the 
simulations in Fig.~4, 5, 6, 7 and 8. Also shown is the 
$\sigma \propto r^{0.5}$ relation, coinciding with most observational 
results. In Fig.~4 and 5, we show the velocity size-scale relation for uniform and
clumpy initial gas distributions, for the shock tube test and the
sinusoidal potential. For the sinusoidal potential, the velocity dispersion for
the uniform shock
dispersion remains flat and subsonic (the sound speed in all simulations is
0.3). This is as expected, since for a uniform shock, the velocity of the
shocked gas should have zero velocity dispersion. 
Again for the shock tube tests, the velocity dispersions for the uniform shocks
are relatively flat.
By contrast the clumpy
shocks show an increasing velocity dispersion with size-scale. 
The velocity of gas in the shock depends on the amount of
mass it has encountered (Section~3.3). For the clumpy shock, gas entering the 
shock will
encounter different amounts of mass (e.g. where gas approaches another clump, or
alternatively a relatively empty area) and a range of velocities are exhibited 
by the shocked gas. This range of velocities increases as the size-scale of the
region increases. At some size-scale, the region of gas will contain the full
range of structure inherent in the initial distribution. The velocity
size-scale relation then remains relatively flat for any further increase in
size-scale.

\begin{figure}
\centerline{\psfig{file=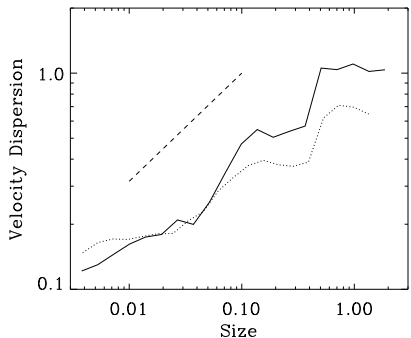,height=2.in}}
\caption{The 1D velocity dispersion dependence on size-scale is plotted
for post-shock gas where the initial distributions are a 2.2D (solid) 
and 2.7 D (dotted) fractal. The gas is passed through a
sinusoidal potential and the maximum spatial extent of the initial fractals is 3. 
The dashed line
show $\sigma \propto r^{0.5}$ and the sound speed is 0.3.}
\end{figure}

Fig.~5 also compares the velocity dispersion relation when constant pressure
boundaries are applied compared to using a diffuse hot medium. The average 
gradient from the 2 slopes is very similar. Although there is some difference 
in the form of the velocity dispersion relation, 
the dynamics appear to be dominated by the cold gas.
Fig.~6 compares the velocity dispersion
calculated using the actual and smoothed velocities, for a uniform and clumpy shock. 
Using the smoothed velocities reduces the velocity dispersion in the uniform shock by a 
factor of around 2. The velocity dispersion is reduced at small scales for the 
clumpy shock leading to a slight increase in the gradient.

In reality, the ISM has structure on many different length scales. To explore
how this affects the resulting velocity dispersion, we have run simulations with
a range of clump sizes and with initially fractal distributions.
Fig.~7 and 8 also show an increasing velocity size-scale
relation for initial gas distributions exhibiting structure on a range of 
scales. Fig.~7 shows the velocity size-scale
dependence where the initial distribution consists of different sized clumps, 
and Fig.~8 the initial fractal distributions. 
The velocity dispersion extends to smaller scales due to the
presence of structure initially over these scales. Again, using the smoothed velocities 
slightly increases the gradient of the velocity dispersion against size-scale.
The velocity size-scale relation is similar to the observed relation
$\sigma \propto r^{0.5}$ for most of the results in Fig.~4, 5, 6, 7 and 8
corresponding to non-uniform initial distributions.
The exponent $\alpha$, where
$\sigma \propto r^{\alpha}$, varies from approximately 0.29 for the 2.7D fractal
distribution to 0.75 for the distribution with different clump sizes. With the
exception of the $\alpha=0.75$ result, our results lie within the observed range of 
values e.g. $\alpha=0.5\pm0.05$ \citep{Solomon1987,Dame1986}, 
$\alpha=0.4\pm0.1$,  $\alpha=0.65\pm0.08$ \citep{Fuller1992},  
$\alpha=0.2,0.5$ \citep{Goodman1998}, representing a range of size scales from 
cloud cores to molecular clouds.

Generally the distributions with higher filling factors produce shallower
gradients in the velocity dispersion relation. As the filling factor increases, 
the distributions and subsequent velocity dispersions
tend towards those of uniform gas, and subsequently the maximum velocity
dispersion is less supersonic. 
The exception appears to be the
distribution of different sized clumps, where the velocity size scale relation 
is somewhat steeper, probably because the filling factor of the smaller clumps
is lower, and the smaller clumps contain most of the mass.

\subsection{Oblique shocks}
Due to the geometry of the shock, the velocity dispersion relations shown in the
previous section have only used the $v_x$ component of the velocity. 
We investigate a more general case by modifying Equation (1) to include a 
dependence on $y$. This produces an oblique shock, more similar to spiral
shocks. The potential is still sinusoidal, 
but the minima of the potential now lie in planes inclined by an angle 
$\theta$ to the $yz$. This is equivalent to an inclination of 
$90^o-\theta$ with the initial flow of the gas.  
We perform a shock test with this modified potential using the clumpy 
($r_{cl}$=0.1) initial distribution.
The gas column density for the shock is shown in Fig.~9. 
The shock is found to induce a velocity dispersion in both $v_x$ and $v_y$ 
although the $v_z$ dispersion is still subsonic.
In Fig.~10, the velocity dispersion is displayed for $v_x$ and $v_y$, where the
shock is inclined at $45^o$ and $30^o$. 
The magnitude of the $v_x$ and $v_y$
dispersions is then proportional to the component of the shock front
perpendicular to these directions, i.e.
\begin{equation}
\frac{\sigma_x}{\sigma_y}=\frac{max(|\sigma| \cos\theta,c_s)}{max(|\sigma| 
\sin\theta,c_s)}
\end{equation} 
for a given size scale. 
The term $c_s$ is an estimate of the minimum value of the dispersion, along the
line of sight parallel to the shock (though as shown next in Fig.~11, the
minimum velocity dispersion in these simulations is slightly lower than $c_s$).

In Fig.~11, we display the velocity dispersion along the line of sight for
different viewing angles, again for the $30^o$ and $45^o$ oblique shocks. 
The maximum velocity dispersion is chosen at each viewing angle, corresponding 
to a size scale of $\thicksim 1$ (c.f. Fig.~10). The angle $\phi$ is measured 
anticlockwise from the $x>0$ axis, with $10^o$ increments. 
The peaks then correspond to the line of sight perpendicular to the shock, 
whilst the minima occur when the line of sight is parallel to the shock.
The difference in magnitude of the peaks occurs since the amplitude of the
$45^o$ shock is stronger.
From these results, we would expect some anisotropy in the magnitude of
the velocity dispersion of gas in molecular clouds, corresponding to the 
geometry of shocks in the gas. For classical models of turbulence, the 
magnitude of the velocity dispersion is unlikely to show any preferred 
direction. 

\begin{figure}
\centerline{
\psfig{file=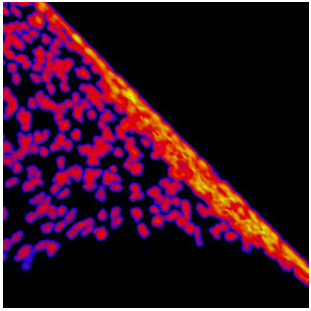,height=1.6in}}
\caption{The column density distribution is shown for an oblique shock where the
shock front is inclined at $45^o$ to the $y$ axis.
The gas initially has a clumpy ($r_{cl}$=0.1) distribution and is passed through a 
sinusoidal potential. The figure ($2.6<x<4.6$, $-0.8<y<1.8$)
shows a stage (t=0.2) during the shock, and the shock in this image lies
half way between a potential minimum and the following maximum. 
The scaling is the same as Figure~1.}
\end{figure}

\begin{figure}
\centerline{\psfig{file=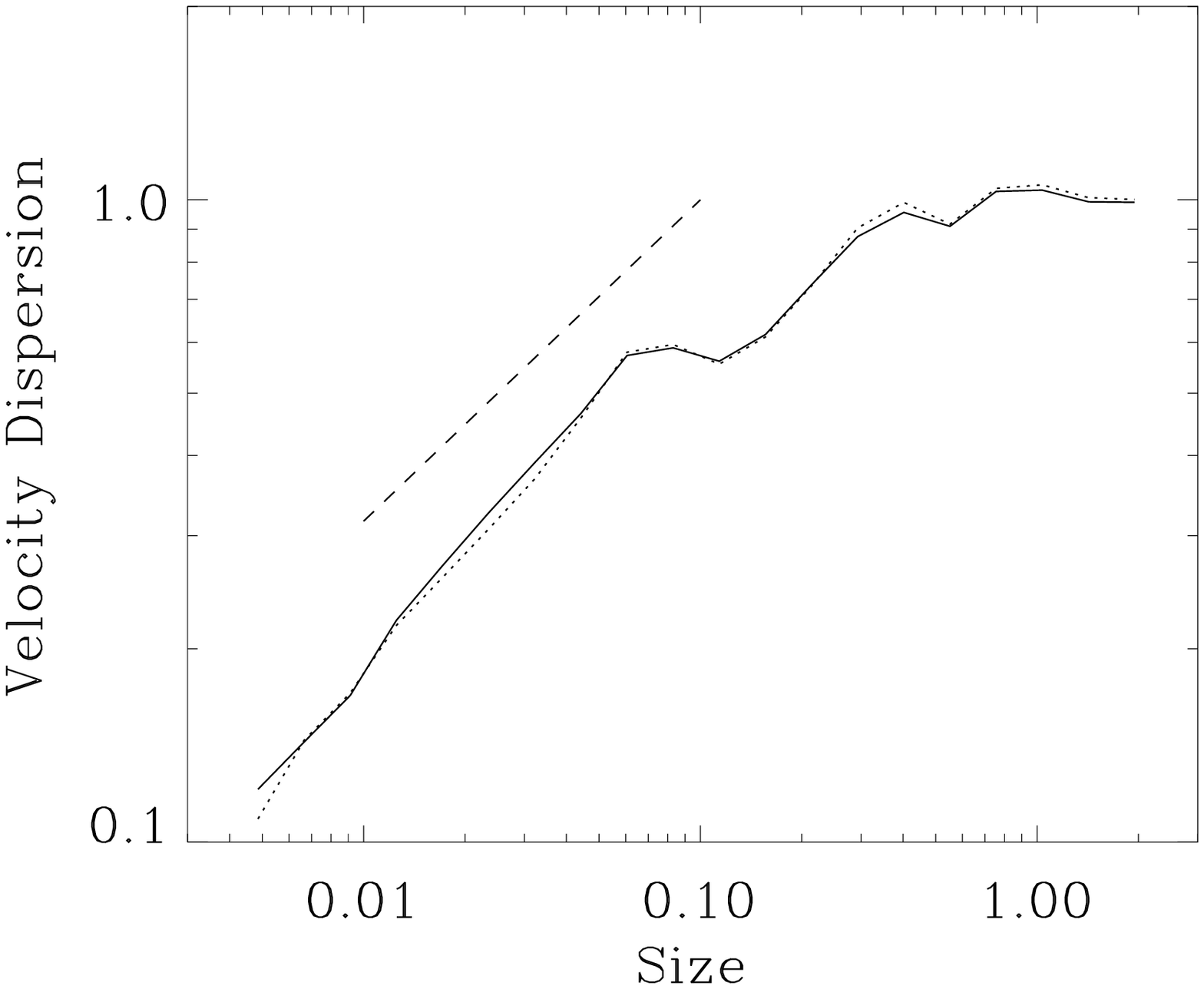,height=1.9in}} 
\centerline{\psfig{file=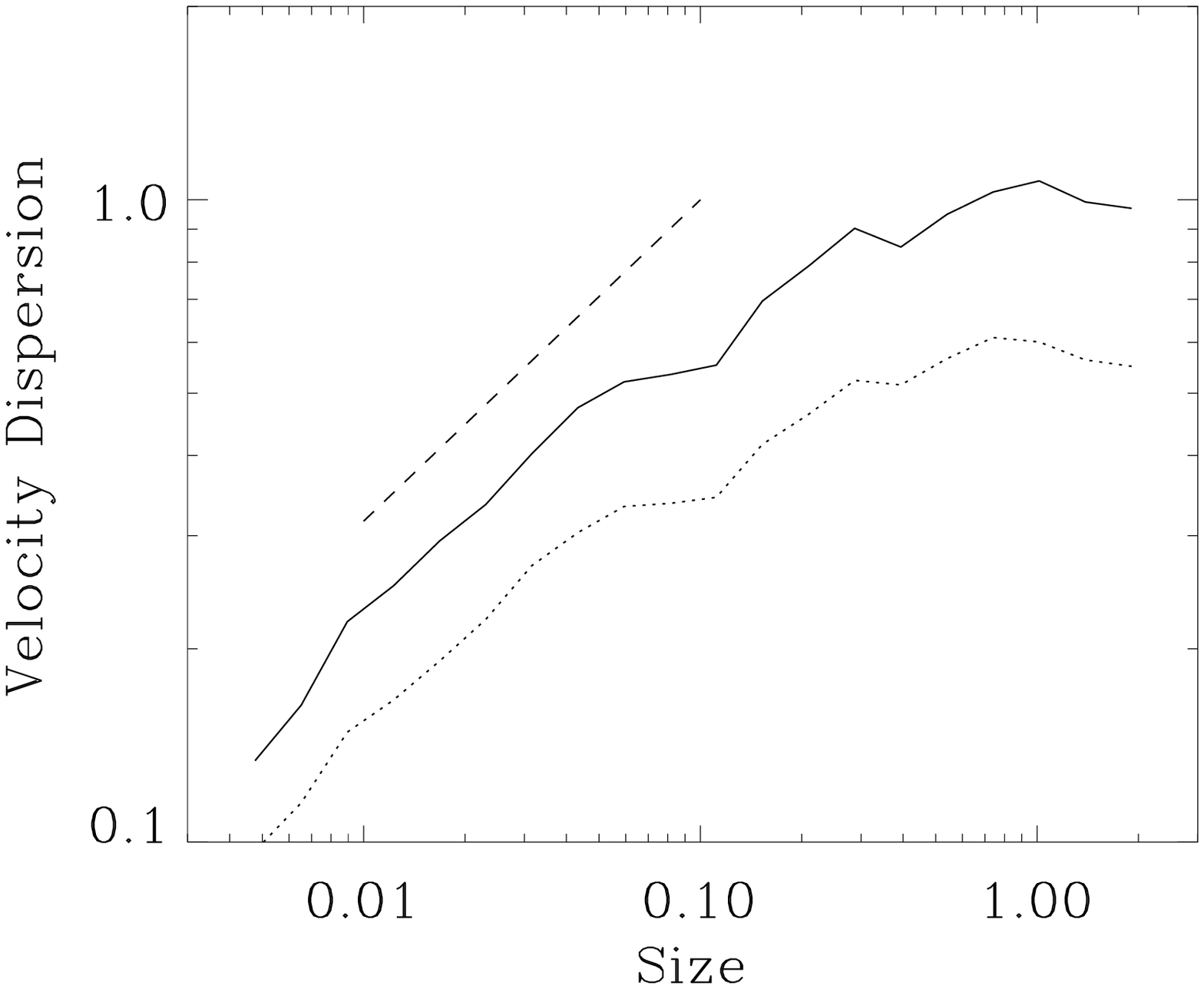,height=1.9in}}
\caption{The $v_x$ (solid) and $v_y$ (dotted) velocity dispersions are plotted
against size-scale for post-shock gas where the shock is inclined at $45^o$
(top) and $30^o$ (bottom) to the y axis. 
For $\theta=45^o$, $\sigma_x$ and $\sigma_y$ are almost identical.
The clumpy ($r_{cl}$=0.1) distribution is
used, with a maximum initial length scale of 3. The gas shocks when passed 
through a sinusoidal potential. The dashed line
shows $\sigma \propto r^{0.5}$ and the sound speed is 0.3.}
\end{figure}

\begin{figure}
\centerline{\psfig{file=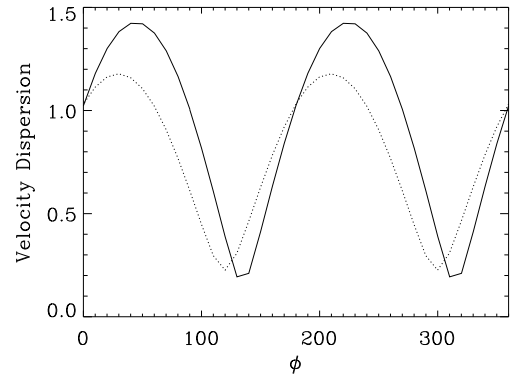,height=2.2in}}
\caption{The maximum velocity dispersion along the line sight is plotted against
viewing angle $\phi$ for the $45^o$ (solid) and $30^o$ (dotted) oblique shocks.
The angle $\phi$ is measured anticlockwise from the $x>0$ axis.}
\end{figure}

\subsection{Mass loading}
We now examine physically why an increasing velocity size-scale relation emerges in our
models. For the clumpy
shocks, the gas exhibits a range of densities and velocities across the region
of shocked gas. The post-shock velocity of a small parcel of gas depends on
the amount of mass it encounters during the shock ('mass loading'). If proportionate amounts of
gas enter the shock, conservation of momentum determines that 
the velocity of the gas in the rest frame of the shock will
be small. However, if gas entering the shock encounters only a small amount of
material, it's velocity will be less affected and remain of higher magnitude.  

Within a region of size-scale less than the structures in the gas, gas in
that region will encounter a similar column density in the shock. Therefore the
gas will exhibit a low velocity dispersion. However as the size-scale of the
region increases, the region will include different structures and gas of
different densities. Therefore different parcels of gas will encounter 
different amounts of
mass and exhibit different post-shock velocities. Thus over a larger region a
higher velocity dispersion occurs in the gas. 
\begin{figure}
\centerline{\psfig{file=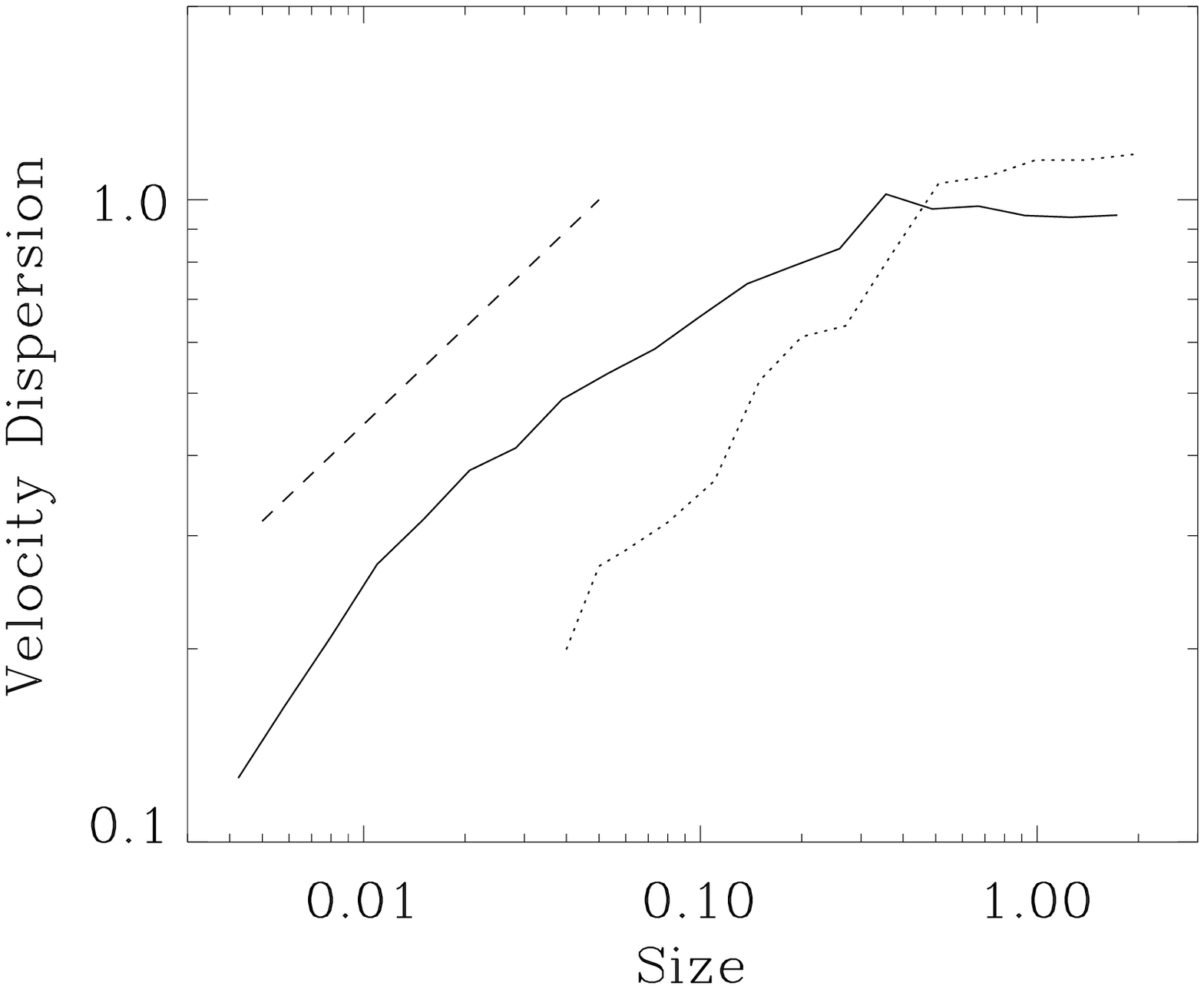,height=2.in}} 
\centerline{\psfig{file=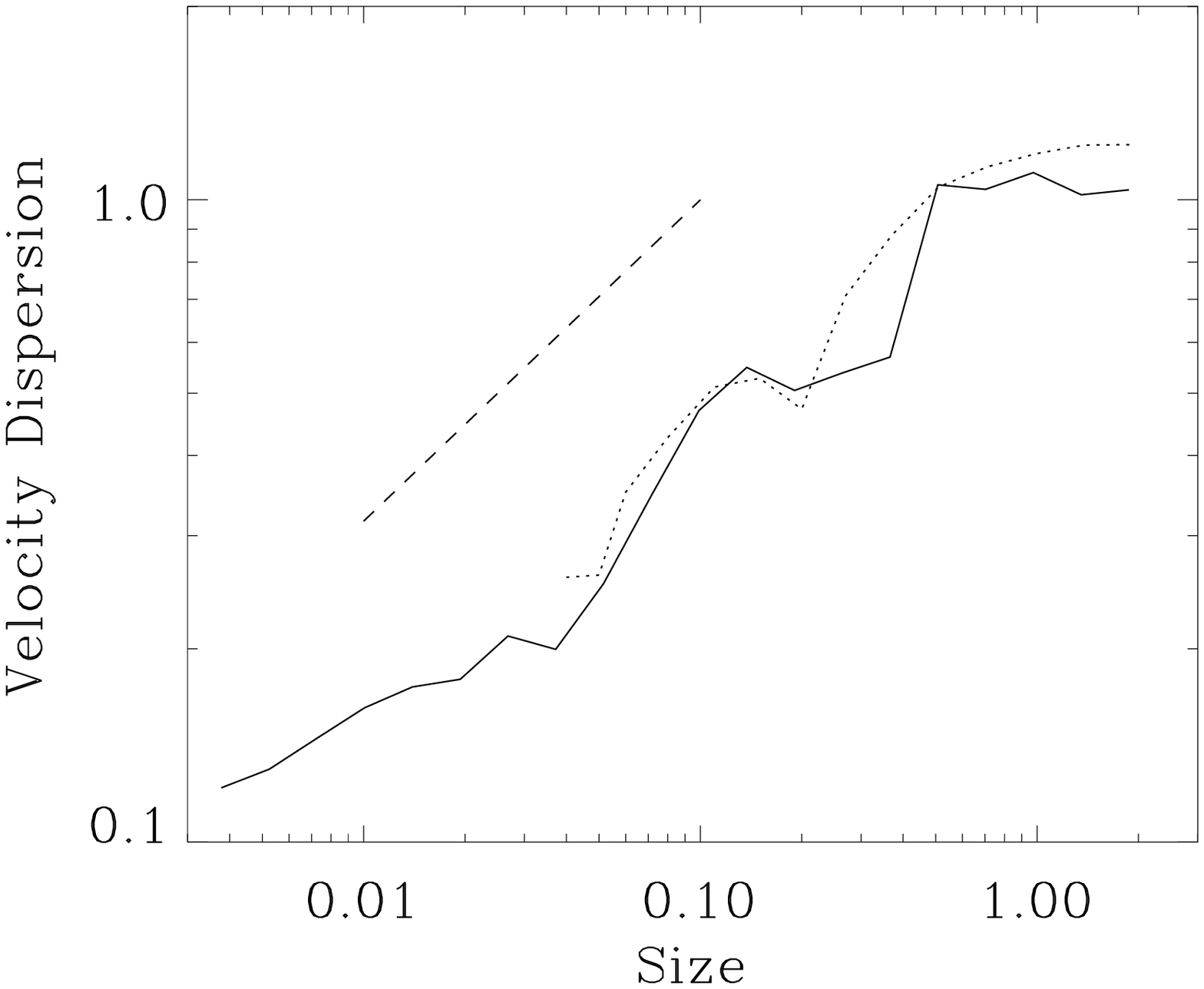,height=2.in}}
\centerline{\psfig{file=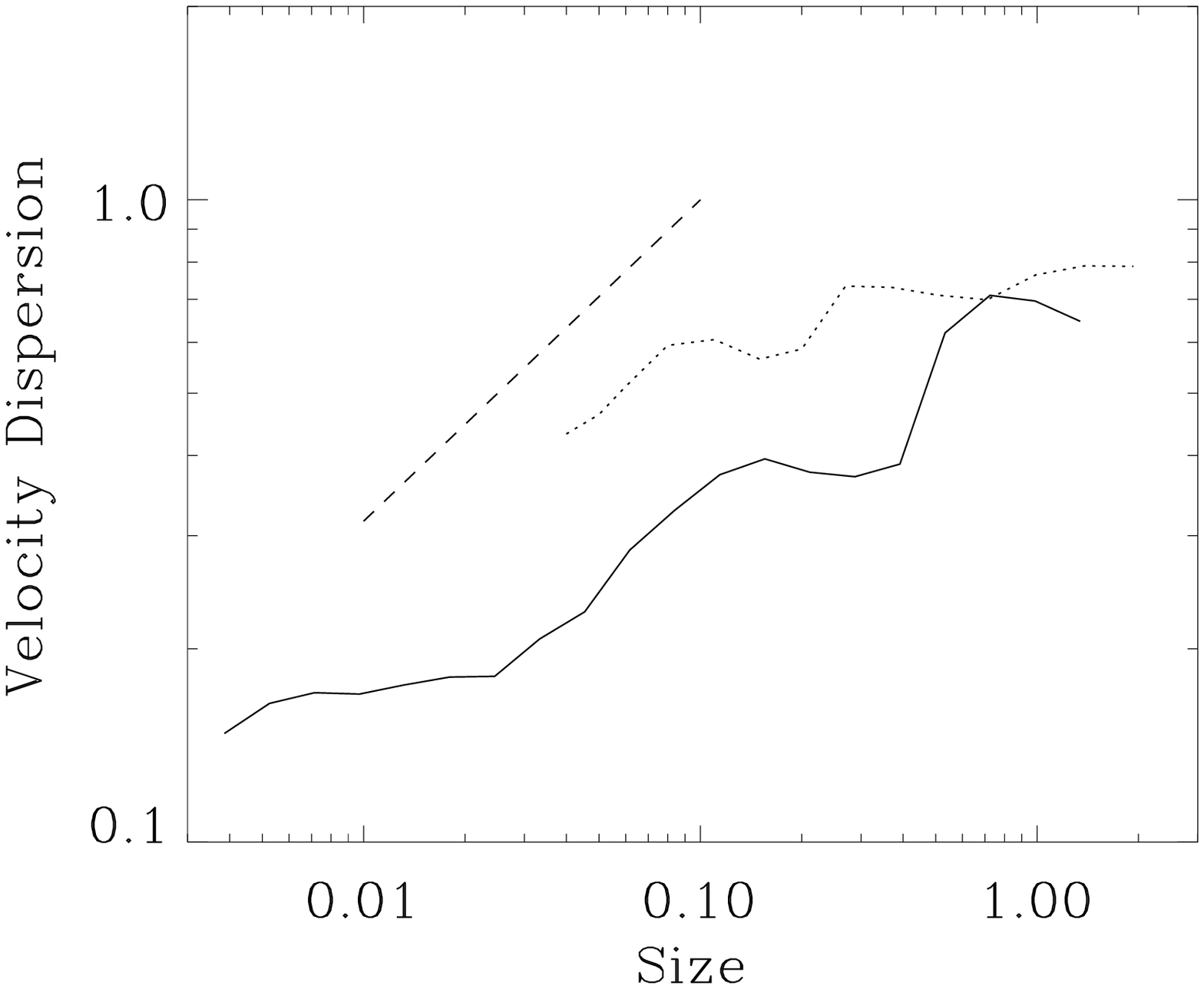,height=2.in}}
\caption{The dependence of the one-dimensional velocity dispersion of the post-shock gas
on size-scale is plotted from our simulations (solid) (as in Fig. 5 and 8)
and the corresponding semi-analytical result (dotted). The semi-analytical result is
calculated assuming the post-shock
velocities are dependent on mass loading. Three different distributions are
shown: clumpy (clump radii 0.1) (top); 2.2D fractal (middle) and 2.7D fractal
(bottom). The dashed line shows $\sigma \propto r^{0.5}$.}
\end{figure}

To test the hypothesis of mass loading, we use a semi-analytical 
approach to calculate the velocity dispersion for the fractal and clumpy 
distributions. We set up the initial distribution of particles
and position a grid across the plane in which the shock will occur, at the
centre of the distribution. 
We then calculate the mass of particles located within a distance $l$  
either side of the grid for each grid 
cell. By applying the conservation of momentum, we determine the expected
velocity of the gas in each grid cell, assuming that all the gas within the
width $l$
of the centre is compressed into the shock. We took $l$=0.2 for the clumpy shock and
$l$=0.6 for the fractal distribution where more of the gas ends up in the shock
(this parameter does not change the results providing sufficient material is
included to accurately represent the gas distribution). 
We then calculate the velocity dispersion
as described in Section~3.1.

In Fig.~13 we display the velocity size-scale relation for the clumpy
distribution (clumps of radius 0.1) and the 2 fractal distributions, from
simulations and the corresponding semi-analytical tests. The results for the 
simulations are all taken from the sinusoidal potential tests. For the clumpy 
distribution the slope determined from mass loading is somewhat steeper than the
results from the simulation. The shape and gradient for 
the 2.2 D fractal shows a 
strong correlation between the simulation and the analytical result.
For the 2.7D fractal, the semi-analytical velocity dispersion size relation is 
shallower than that determined by the simulation, although both show a shallower
gradient compared to the other 2 distributions. 
The analytical method can be repeated by setting up distributions with different
random seeds to show the degree of scatter in the expected gradients for the
distributions. The clumpy shock, which should give very similar distributions
regardless of the initial seed, produced a consistent slope of 
$\alpha=0.48\pm0.03$ compared to $\alpha \approx 0.43$
for the simulation. The fractal distributions showed a much greater variation in
slope, the 2.2D fractal giving $\alpha=0.43\pm0.16$ compared to $\alpha \approx 
0.4$ for the simulation.   
For the 2.7D fractal, we found $\alpha=0.25\pm0.15$ for the analytical result,
so the line shown on Fig.~12 is at the lowest extent of this range, 
whilst $\alpha \approx 0.3$ for the simulation.  

\section{Analytical models}

\subsection{Collision of two clumps}
The previous results represented plausible mass distributions for the ISM,
although consequently we were only able to calculate the velocity size-scale 
relation numerically. Here we consider the case of 2 clumps colliding and
calculate directly the resulting velocity dispersion.
This distribution represents the 'least uniform' distribution, so the most
removed scenario from a uniform shock. 
The clumps are assumed to be uniform density spheres and are
travelling with equal and opposite velocities.
The clumps are offset from each other and so do not collide head on.
The parameters describing the collision are shown in Fig.~13, where $b$ 
is the impact parameter of the 2 clumps. In the following description, we choose
a Cartesian axis centred on the point of maximum overlap between the 2 spheres,
such that the centre of each sphere is situated at $y=\pm b/2$.
In these coordinates, the collision of the clumps is 
symmetrical about $z=0$.
We take $v_0=1$, and the column density is scaled so that
maximum column density is $\Sigma_{max}=1$ along the diameter of each sphere.
We first assume that the radii of the clumps are the same, $r_1=1$ and $r_2=1$.

To calculate the velocity dispersion, we assume that all the gas in the 2
clumps is compressed when they collide. We calculate the post-shock
velocity of the material by determining, for each clump, the mass per unit area 
perpendicular to the shock ($\Sigma1$ and $\Sigma2$ in Fig.~13). 
The calculation of the velocity dispersion is centred along the column
which contains the most mass. In Fig. 13, this corresponds to the column of
gas at $y=0$, $z=0$ (i.e. the midpoint of $b$).  

In Fig.~14 we plot the column density for each clump and the post-shock velocity, for a 
cross-section of the collision along the $y$ axis.
By taking a
cross-section, the problem is reduced to 1 dimension, and the mass for each
sphere can be determined as a function of length scale. 
Using the idea of mass loading we assume that all gas is compressed into the
$xy$ plane.
We calculate the expected velocity of gas in the shock by applying 
the conservation of momentum, i.e.
\begin{equation} 
v_s=\frac{(\Sigma1-\Sigma2) v_0}{(\Sigma1+\Sigma2)} .
\end{equation} 
In Fig.~14, $b=0.5$ and the 2 spheres are overlapping by a distance of 1.5.
At $y=0$ there is an equal contribution in mass from each sphere, and so the 
velocity in the shock will be 0. 
The velocity increases to $\pm1$ for gas which does not encounter any
material from the other clump.

\begin{figure}
\centerline{\psfig{file=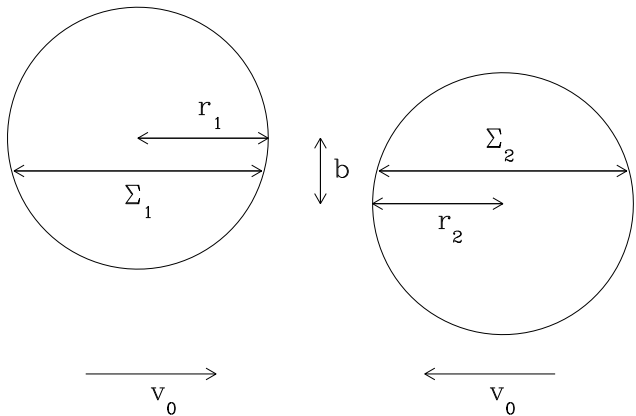,height=2.6in}}
\caption{Diagram showing a collision with impact parameter $b$ 
between 2 spheres. The column densities $\Sigma1$ and $\Sigma2$ 
perpendicular to the shock are calculated to determine the expected velocity 
of gas in the shock.}
\end{figure}

\begin{figure}
\centerline{\psfig{file=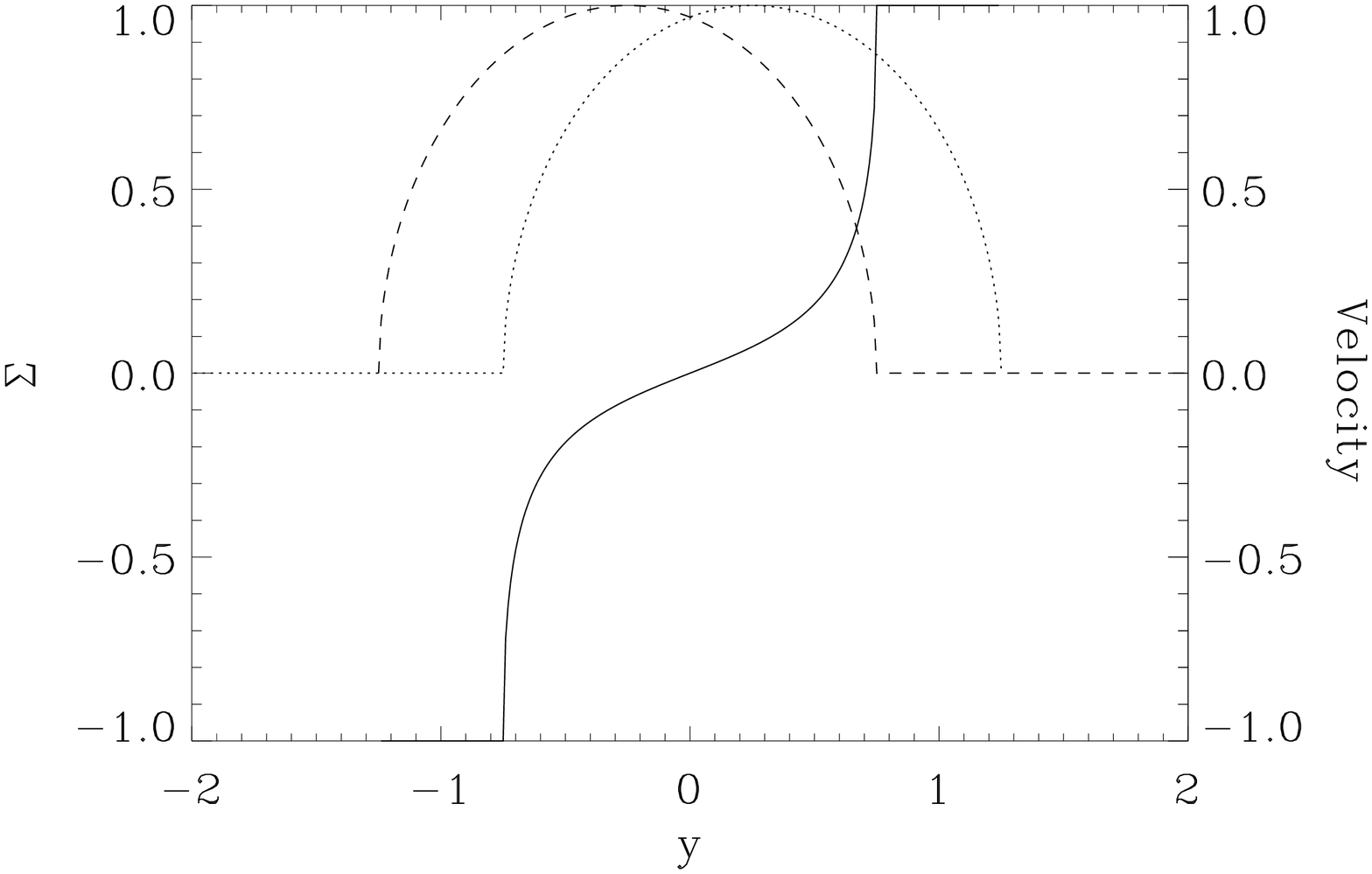,height=2.in}}
\caption{The velocity (solid line) is plotted for a cross-section of
post-shock material when 2 spheres collide. The cross-section corresponds to
taking a cut where $z=0$.
The dotted and dashed lines show
the column density for each sphere along the cross-section.}
\end{figure}

\begin{figure}
\centerline{\psfig{file=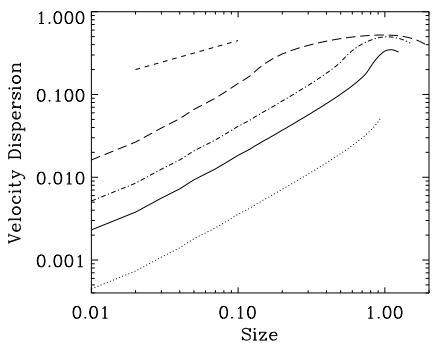,height=2.in}}
\caption{The mass-weighted velocity dispersion predicted for the collision of 2 spherical
clumps, each of radii 1 and initially travelling at $v_0=1$. The impact
parameters of the collision are $b=0.1$ (dotted), $b=0.5$ (solid), $b=1$
(dot-dash) and $b=1.75$ (long dash) whilst the 
short dash line represents $\sigma \propto r^{0.5}$ .}
\end{figure}

\begin{figure}
\centerline{\psfig{file=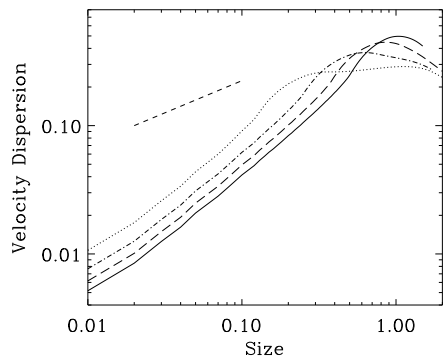,height=2.1in}}
\caption{The mass-weighted velocity dispersion predicted for the collision of 2 spherical
clumps travelling with initial velocity $v_0=1$. The radius of the first clump 
is set to 1, whilst the radius of the second clump is 1 (solid), 0.75 (long
dashed), 0.5 (dot-dash) and 0.25 (dotted). The impact parameter for all the
collisions is $b=1$. The short dash line shows $\sigma \propto r^{0.5}$ .}
\end{figure}

We extend this idea to calculate the velocity across a 2D shock from the
collision of the clumps. We again assume that the gas is compressed in the $yz$
plane and calculate the mass density entering the shock 
from each sphere. We then determine the expected
post-shock velocity of the gas, again over the $yz$ plane. We
calculate the velocity dispersion over a disk centred on the
co-ordinates $y=0$, $z=0$. The velocity
dispersion is determined over disks of increasing radius, to produce a velocity
size-scale relation. We plot the mass-weighted velocity dispersion in Fig.~15, 
with collisions of different impact parameter.

For the simple case of 2 clumps colliding, the velocity dispersion increases
with size scale approximately as a power law, independent of the impact 
parameter. Increasing the impact parameter transposes the power law to smaller
size scales, since the size of the shocked region
decreases. For size-scales $>r-b/2$, the velocity
dispersion includes gas which does not enter the shock and the velocity
size-scale relation flattens.
The power law for the shocked gas is somewhat steeper than the
$\sigma \propto r^{0.5}$ relation observed, rather $\sigma~\propto~r^1$.     
However, the power law exponent may be expected to 
decrease when considering larger regions and structure on
multiple scales composed of many clumps (Section~4.2).

This power law is also independent of the comparative radii of the 2
spheres. Fig.~16 shows the effect of
varying the radius of the second clump ($r_{2}$). 
Similarly to varying the impact parameter, the velocity size-scale relation 
shifts to smaller scales as
$r_{2}$ decreases, and levels off at size-scales comparable to 
$r_{2}$. Scaling both clump radii up or down will extend the velocity
size-scale relation to larger or smaller size-scales. For example, if the radius
decreases by a factor of 10, the column density and therefore mass-weighted
velocity dispersion will also decrease by the same factor.  

Assuming that the clumps have velocities $v_0=1$, the velocity
dispersion for an impact parameter of $b\ge 0.5$ reaches a maximum of
$\thicksim$ 0.3. 
Since the initial velocity $v$ can be scaled
up or down, the maximum velocity dispersion for an initial velocity of 
$v_0 c_s$ will be approximately 0.3 $v_0 c_s$ (e.g. 3 $c_s$ if 
${\cal M}=20$). When the impact parameter is small, the velocity
dispersion is unlikely to be supersonic (e.g. the maximum is $\thicksim 0.5$ 
$c_s$ when $b=0.1$ and ${\cal M}=20$). This is expected, since if 
$b=0$ the clumps collide head on and the velocity dispersion is 0 everywhere.  

\begin{figure}
\centerline{\psfig{file=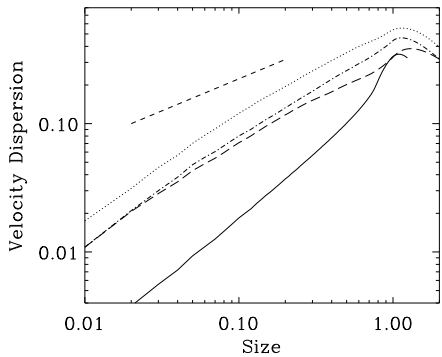,height=2.1in}}
\caption{The mass-weighted velocity dispersion predicted for the collision of
multiple spherical clumps. The solid line shows the velocity dispersion for 
2 clumps of radii 1 and impact parameter 0.5. The dot-dash and dotted lines 
give the average velocity dispersion for multiple collisions of clumps, where 
all clumps are assumed to have radii of 1, but the impact parameter varies. 
The impact
parameter is either random (dot-dash) or biased towards large $b$ (dotted).
The velocity dispersion for multiple collisions with both random $b$ and 
$r_2$ is given by the long dashed line. The short dash line shows
$\sigma \propto r^{0.5}$.}
\end{figure}

\subsection{Multiple collisions of clumps}
The velocity size-scale relation obtained for 2 clumps colliding is somewhat 
steeper than those observed for molecular clouds. 
However this distribution of gas is an extreme case, as in general the gas would
be comprised of multiple clumps of various sizes. 
We now consider many collisions and calculate the average velocity dispersion
at each size scale.
In all the collisions, we still assume 2 clumps collide, and fix
the radius of the first clump to 1. However both the impact parameter $b$ and
the second clump radius, $r_{2}$, can be varied. 
We show 3 different possibilities for the collisions of clumps 
in Fig.~17. Firstly we vary $b$ randomly, assuming a probability
distribution function of 
$f(b)=1$, $0<b<1$, with $r_2$ fixed at 1. Secondly, we take $b$ biased toward
larger values, so $f(b)=b$, $0<b<1$, and $r_2$ is again fixed at 1. 
Finally, we allow both $b$ and
$r_2$ to vary randomly, so $f(b)=1$, $0<b<1$ and $f(r_2)=1$, $0<r_2<1$.
For comparison, the velocity size-scale relation from 2 clumps colliding with 
impact parameter $b=0.5$ (and both of radii 1) is also included. 

The slope is 
shallower for multiple collisions of random $b$, giving $\sigma
\propto r^{0.8}$. 
The velocity dispersion was also calculated with $b$ biased towards large
values, reflecting the probability of a collision relative to
the impact parameter. This produces a slightly flatter slope, although it mainly
increases the magnitude of the velocity dispersion.
Finally, with both $b$ and $r_{2}$ allowed to vary randomly, the
gradient decreases further to $\sigma \propto r^{0.7}$. 

This is still an incomplete model compared to the simulations of clumpy gas 
previously described, with further possible variations in the clump and shock
geometry. Furthermore, collisions of multiple clumps have not been included, 
i.e. where $>$ 2 clumps collide simultaneously, or 2 clumps collide with each 
other before colliding with further material in the shock. The latter is 
evident in the simulations, where layers of clumps enter the shock and interact with
each other. This is again likely to produce a shallower relation compared with 
the calculations in this section.
Overall, for the most structured distribution of 2 offset clumps colliding, the
resulting velocity size-scale relation is steep, $\sigma \propto r^1$. For the 
least structured distribution, i.e. uniform gas, the velocity size-scale 
relation of the shock is flat. In reality, the distribution of gas, and therefore
the gradient of the velocity size-scale relation, is likely to lie within these 
two extremes.   

\section{Conclusion}
We have presented simulations and analysis of shocks for initial distributions
of uniform, clumpy and fractal gas.
We find an increasing velocity size scale relation in all our results similar to
that observed, except for the uniform shocks.
For example, for a 2.2 D fractal distribution representative of the ISM, the 
velocity size-scale relation is approximately 
$\sigma \propto r^{0.4}$, in good agreement with observations.
The slope of the velocity size-scale
relation tends to decrease with distributions corresponding to higher filling 
factors.  
The velocity size-scale relations determined for these distributions can be
understood in terms of mass loading, as indicated by our analytical tests.
Oblique shock tests show that the magnitude of the line-of sight velocity 
dispersion depends on the angle from which the shock is viewed, 
providing an observational test for this model. 

These results imply that: 1)  
The observed multi-scale structure of the 
ISM may explain the velocity dispersion in molecular clouds. This is in contrast
to the usual view that turbulence produces the structure of molecular clouds 
(e.g. \citet{Falgarone2005});
2) In these
models, the dynamics of the shocked gas corresponds to random velocities, rather than 
classical turbulence, despite the apparent velocity size scaling relation.

\section*{Acknowledgements}
We thank the referee for useful comments and suggestions.
Computations included in this paper were performed using the UK Astrophysical
Fluids Facility (UKAFF).
   
\bibliographystyle{mn2e}
\bibliography{Dobbs}

\bsp

\label{lastpage}

\end{document}